\documentclass[]{birkjour}
\usepackage[noadjust]{cite}
\RequirePackage[all]{xy}

\usepackage[utf8]{inputenc} 
\usepackage[T1]{fontenc}    
\usepackage{hyperref}       
\usepackage{url}            
\usepackage{booktabs}       
\usepackage{amsfonts}       
\usepackage{nicefrac}       
\usepackage{microtype}      
\usepackage{multirow}
\usepackage{amsmath}
\usepackage[caption=false]{subfig}

\usepackage{graphicx}

\usepackage{cite}

\usepackage[dvipsnames]{xcolor}

\newtheorem{lemma}{Lemma}

\usepackage{algorithm,algcompatible,amsmath}
\algnewcommand\INPUT{\item[\textbf{Input:}]}%
\algnewcommand\OUTPUT{\item[\textbf{Output:}]}%

\usepackage{algpseudocode}

\algnewcommand\algorithmicforeach{\textbf{for each}}
\algdef{S}[FOR]{ForEach}[1]{\algorithmicforeach\ #1\ \algorithmicdo}

\begin{document}

\title[Efficient Networked VR Transformation Handling Using GA]{Less Is More: Efficient Networked VR Transformation Handling Using Geometric Algebra}

\author[Kamarianakis]{Manos Kamarianakis}

\address{%
Department of Mathematics \& Applied Mathematics, \\
University of Crete, \\
Voutes Campus, 70013 Heraklion, Greece\\
Orchid ID: 0000-0001-6577-0354}

\email{kamarianakis@uoc.gr}

\thanks{The authors are affiliated with the University of Crete, Greece 
and the ORamaVR company\cite{oramavr}.
This is an extended version of work originally presented in the CGI 
2021 conference, on the ENGAGE workshop 
\cite{kamarianakisNeverDropBall2021}. 
Corresponding Author: \url{kamarianakis@uoc.gr}}

\author[Chrysovergis]{Ilias Chrysovergis}
\address{ORamaVR, \\
FoRTH (STEP-C), 70013 Heraklion, Greece\\
Orchid ID: 0000-0002-5434-2175}
\email{ilias@oramavr.com}

\author[Lydatakis]%
{Nick Lydatakis}
\address{Department of Computer Science, \\
University of Crete, \\
Voutes Campus, 70013 Heraklion, Greece\\
Orchid ID: 0000-0001-8159-9956}
\email{nick@oramavr.com}

\author[Kentros]{Mike Kentros}
\address{Department of Computer Science, \\
University of Crete, \\
Voutes Campus, 70013 Heraklion, Greece\\
Orchid ID: 0000-0002-3461-1657}
\email{mike@oramavr.com}

\author[Papagiannakis]{George Papagiannakis}  
\address{Department of Computer Science, \\
University of Crete, \\
Voutes Campus, 70013 Heraklion, Greece\\
Orchid ID: 0000-0002-2977-9850}
\email{papagian@ics.forth.gr}



\begin{abstract}
As shared, collaborative, networked, virtual environments become 
increasingly popular, 
various challenges arise regarding the efficient transmission 
of model and scene transformation data over the network. 
As user immersion and real-time interactions heavily 
depend on VR stream synchronization, transmitting the entire 
data set does not seem a suitable approach, 
especially for sessions involving a large number of users. 
Session recording is another momentum-gaining feature of 
VR applications that also 
faces the same challenge.
The selection of a suitable data format can reduce the occupied volume,
while it may also allow effective replication of the VR session 
and optimized post-processing for analytics and deep-learning algorithms.

In this work, we propose two algorithms that can be applied 
in the context of a networked multiplayer VR session,
to efficiently transmit the displacement and orientation data 
from the users' hand-based VR HMDs. 
Moreover, we present a novel method for effective VR recording
of the data exchanged in such a session. 
Our algorithms, based on the use of 
dual-quaternions and multivectors, impact the 
network consumption rate and are highly effective in 
scenarios involving multiple users. By sending less data over the 
network and interpolating the in-between frames locally, we manage to 
obtain better visual results than current state-of-the-art 
methods. Lastly, we prove that, for recording purposes,
storing less data and interpolating them on-demand yields a data set 
quantitatively close to the original one.
\end{abstract}

\maketitle

\begin{figure}[htbp]
  \centering
  \subfloat[]{\includegraphics[width=0.32\textwidth]{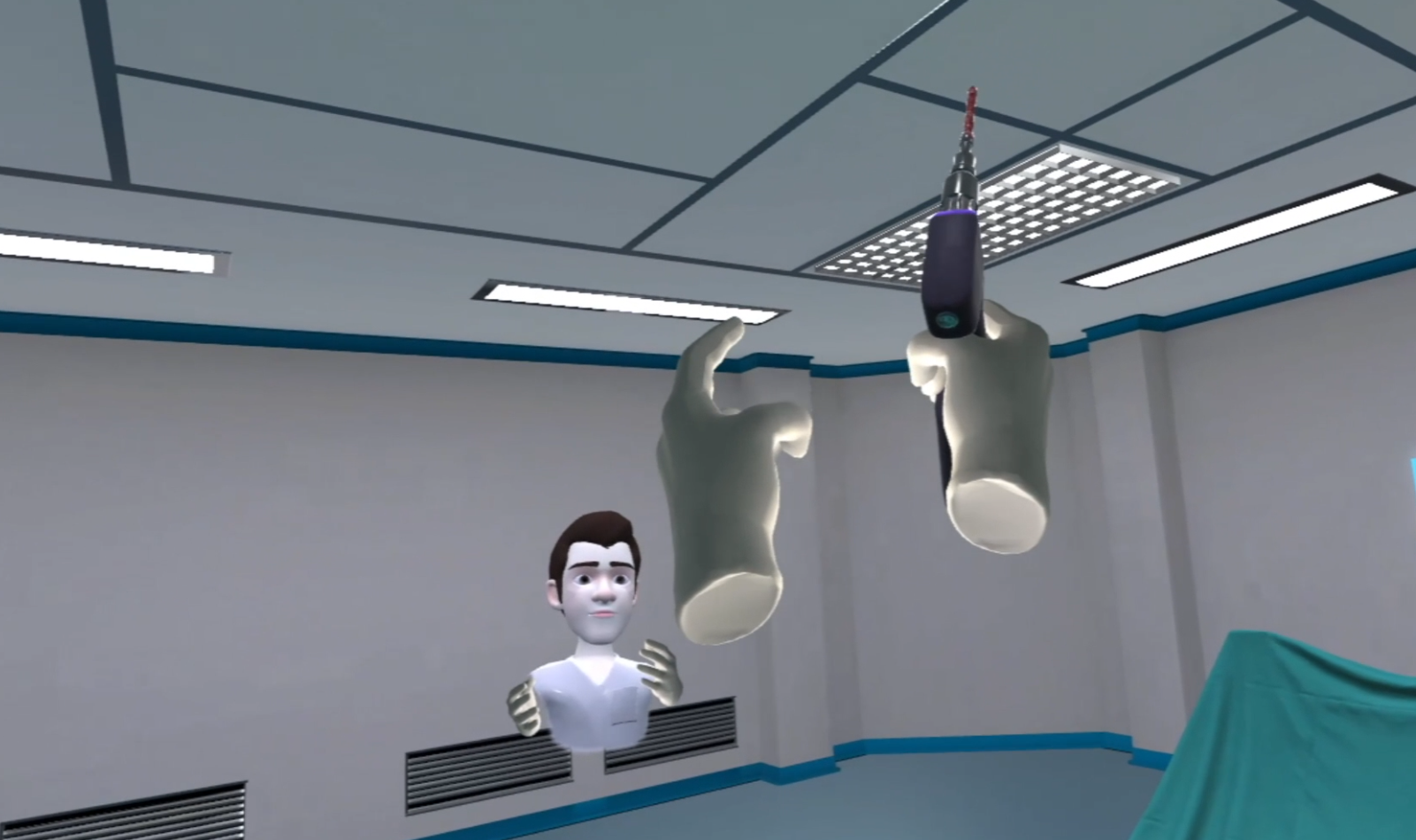}}%
  \hspace{3pt}%
  \subfloat[]{\includegraphics[width=0.32\textwidth]{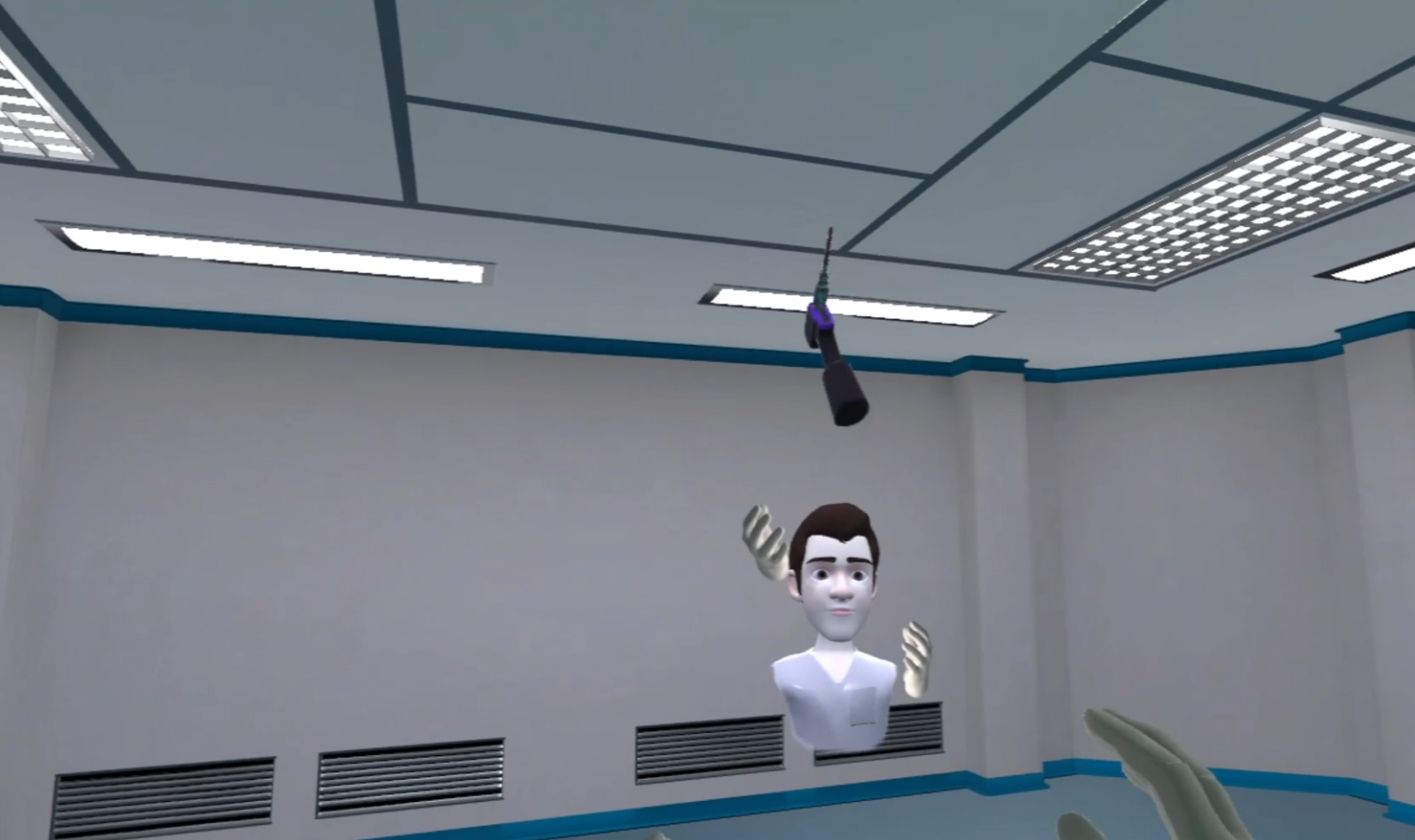}}%
  \hspace{3pt}%
  \subfloat[]{\includegraphics[width=0.324\textwidth]{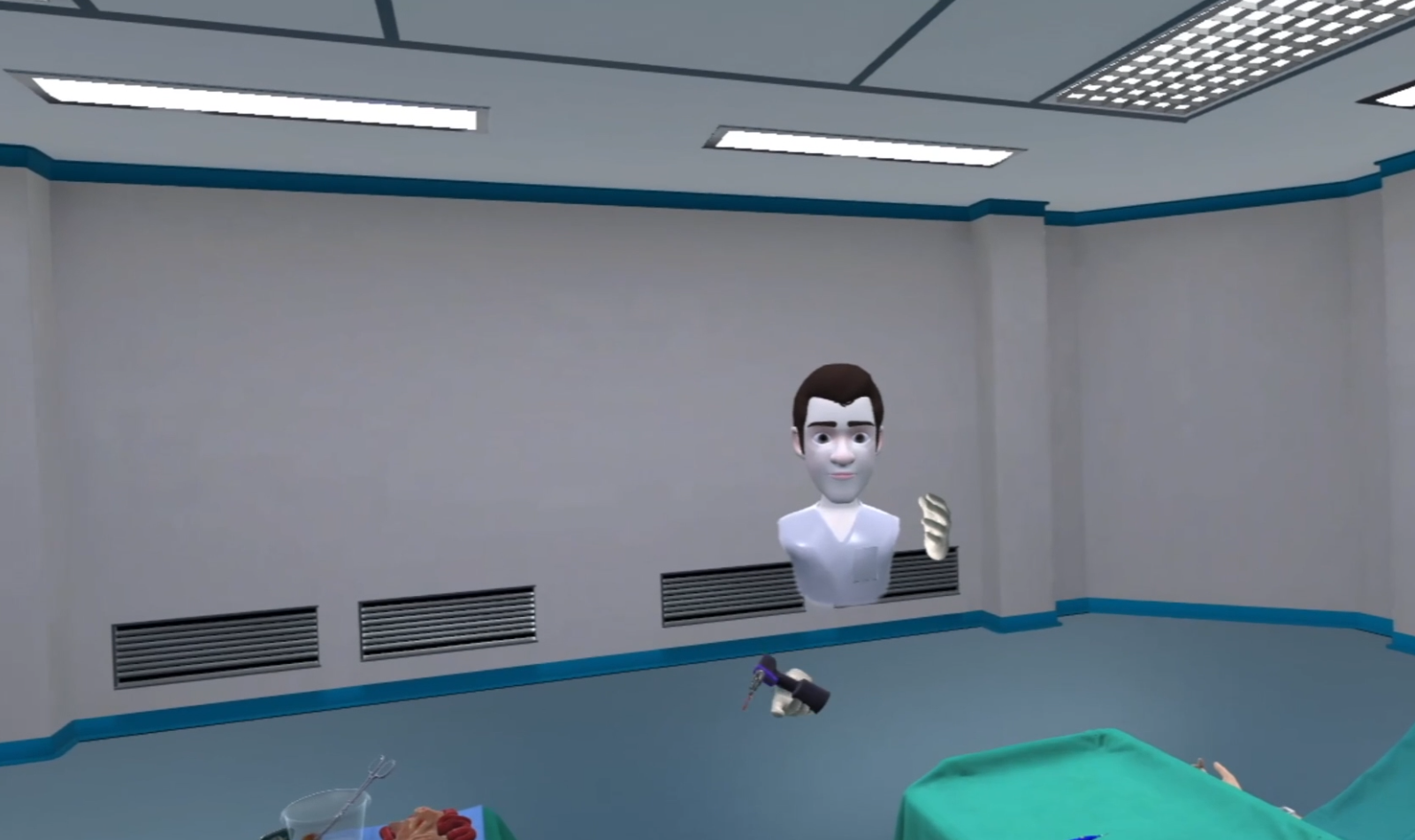}}%
  \caption{Catching a tool in a VR collaborative scenario. 
  (a) A user throws a tool (in our case a medical drill) at another. 
  (b) The object's keyframes, sent by the user  
  that threw it, are interpolated using multivector LERP 
  (see Section~\ref{sub:proposed_method_based_on_multivectors}) 
  on the receiver's VR engine. (c) The receiver manages to 
  catch the tool, as a result of the effective frame generation 
  that is visualized in his/her HMD. }
  \label{fig:catching}
\end{figure}

\section{Introduction} 
\label{sec:introduction}

The rise of the 5G networks and their augmented capabilities, 
along with the increasing need for remote working and 
collaboration, due to the extended pandemic, have influenced 
the research initiatives in many scientific areas.
Virtual Reality (VR) is an area drastically affected by these 
facts.  
Rapid technological advancements of the past decades, both on 
hardware and software level, have enabled VR experts to 
deliver powerful visualization algorithms, optimized to provide 
an immersive user experience. 
VR applications that exploit modern GPU capabilities enable 
the simulation of high-fidelity content of a great variety. 
Driven by modern needs and expectations, collaborative, shared 
virtual environments (CVEs)  have been, upon their inception, 
a highly active research topic 
\cite{churchill1998collaborative,molet1999anyone,
papagiannakis2008survey,ruan2021networked}. 
One of the main research objectives regards the immersion
of the users that are present and interact within the same 
virtual environment. 
Such a context is common and dictated by modern era industries 
that require frequent and efficient remote communication 
between people.
As safety has become a first priority due to the pandemic, 
VR has been proven to be a solution that natively addresses many 
of the challenges posed, while some others require extra effort 
to be effectively resolved.

One such interesting problem that arises within the context of a 
VR CVE, is the provisioning of users with the same high quality of 
experience (QoE) in both single- and multi-player VR sessions. 
As modern Head-mounted Displays (HMDs) 
are able to support intensive resource-demanding VR applications, 
we can deliver high-fidelity and immersive content 
in single user scenarios, even when the user interacts in 
real-time with the virtual environment.
Such experience cannot be natively and/or efficiently 
replicated in the context of multi-user scenarios, even by 
modern game engines. 
The basic challenge that VR application 
designers have to overcome 
regards the exchange of information between the user movements 
and their actions in real-time. 
If this information is not efficiently transmitted, the virtual 
avatars of the users will not be or appear to be in sync. 
In such cases, this unnatural-feeling caused by the 
delay between the users actual movements/actions and their 
visualization, drastically 
reduces the immersion and,  
in certain cases, may even make the application incapable to 
serve its cause, e.g., in VR training applications. 

Another interesting problem that experts try to handle deals with 
session recording. 
Nowadays, session recording and playback of a single or multi-user VR 
session has become of increasing importance for the
functionality and effectiveness of certain applications.
This is especially significant for applications related to 
training as replaying users actions can serve as an additional
and powerful educational tool. 
As users are allowed to watch and study all actions of a virtual scene, they can identify and learn from their mistakes. 

Recording (and replaying) a VR session is not a task that is natively 
undertaken by modern game engines and therefore most VR applications 
do not include such a feature by default. 
Researchers try to determine the optimal way to record and store 
all events that happen in a VR session where a single or multiple 
users exist and interact. 
Their goal is to explore the proper methods and structures of data that 
must be employed so that they can achieve a real-time logging where the 
required data storage remains manageable. 
Ultimately, such recordings are used to enable replaying of a VR session, 
a feature that many VR applications still lack today. 

Current bibliography contains numerous examples of how the VR record and 
replay features can enhance the functionality of a VR application, 
especially the ones related to training, by mainly measuring the 
performance of users. In \cite{Birrenbach2021,Hooper2019}, the authors present  
randomized controlled trials on the efficacy of VR simulation for 
medical skills training. 
In both these works, the VR cohort 
demonstrates greater improvement in each specific score category and 
significantly higher satisfaction, 
compared to the control group.
Grantcharov et al. \cite{Grantcharov2004} demonstrate in their 
randomized clinical trial that surgeons who received VR training 
performed laparoscopic cholecystecomy faster than 
the control group, showing great improvement 
in error and economy of movement scores.
Ihemedu-Steinke et al. \cite{Lhemedu2018} measured the concentration, involvement and enjoyment of users on a virtual automated driving simulator and provided a comparison of those between a VR and a conventional display. 
The experiment showed a statistically significant result, that in VR all measured variables had higher values. 
Southgate \cite{Southgate2020} proposed the use of screen capture video to understand the learning outcomes in virtual reality experiences. 
Paying attention to attitudinal change, higher order thinking and metacognition, it was shown that this was a viable method for gaining deeper insights into immersive learning. 
Lahanas et al. \cite{Lahanas2015} presented a novel augmented reality simulator for skills assessment in minimal invasive surgery. 
Their simulator allows training of surgeons and assessment of their skills, 
while the completion time, the path length and two specific types of errors are evaluated. 

In \cite{Kloiber2020}, analysis of the motion of virtual reality users is presented. 
The authors use the motion captured in virtual environments and perform the analysis in the same environment. 
A visual analysis system is developed that allows immersive visualization of human motion data. 
This method enables the examination of behaviours and can find useful patterns and outliers in sessions.

In \cite{Ninsky2015}, kinematic analysis of experienced and novice surgeons during robotic teleoperation was performed. 
The same transformation data with our methods were used in order to assess the skills of the users. 
However, robotic kits are not capable of fast prototyping of realistic environments and also event data need to be captured from a video camera. 
Also, replay is only available through cameras and has the drawbacks that are explained in Section~\ref{sec:VR_Recorder}. 

Sharma et al. \cite{Sharma2014} presented a framework for video-based objective structured assessment of technical skills. 
In this framework, motion feature extraction is performed to detect the spatio-temporal interest points (STIPs). 
After that, the motion classes are learned using $k$-means clustering and the STIPs are classified into those classes. 
Finally, motion class counts, data-driven time windows and sequential motion texture features are computed.  
An automated video-based assessment tool for surgical skills was proposed in \cite{Zia2016}. 
In this work, a four-step process is developed in order to evaluate the training outcomes in medical schools. 
These steps are motion class time series generation, feature modelling, feature selection, and classification. 
In \cite{Zia2018}, accelerometer data are used in conjuction with video, providing features from multi-modal data. 
This fusion of video and acceleration features can improve performance for skill assessment. 
The video data are processed using techniques similar to the two previous papers, while accelerometer data processing is performed by aligning those data with the timeseries computed by the video data. 
Afterwards, these two kinds of data are combined in order to extract and select useful features and skills assessments are classified using a nearest neighbour classifier.

It becomes apparent from these works and the current research initiatives 
that the use of VR simulations with the capability of VR recording and 
replaying, will enhance the learning outcomes of the users, since this 
functionality enables the replay of sessions and, also, provides data 
to analyze, with automated tools, the performance of the users. 
The latter empowers serious game developers to create assessment tools 
for large-scale real-time evaluations without the need of human labor.

\begin{figure}[htbp]
  \centering
  \includegraphics[width=0.31\textwidth]{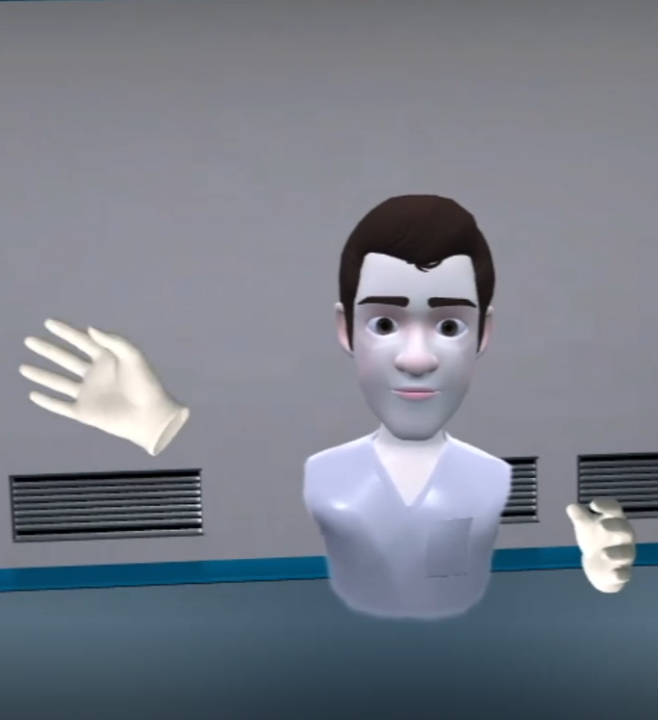} 
  \includegraphics[width=0.33\textwidth]{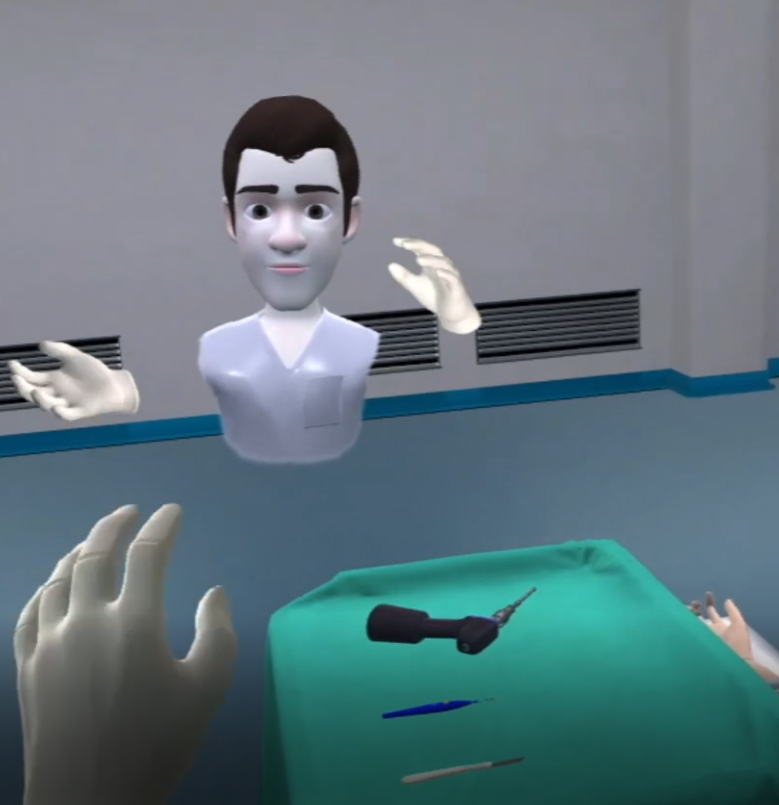}
  \includegraphics[width=0.31\textwidth]{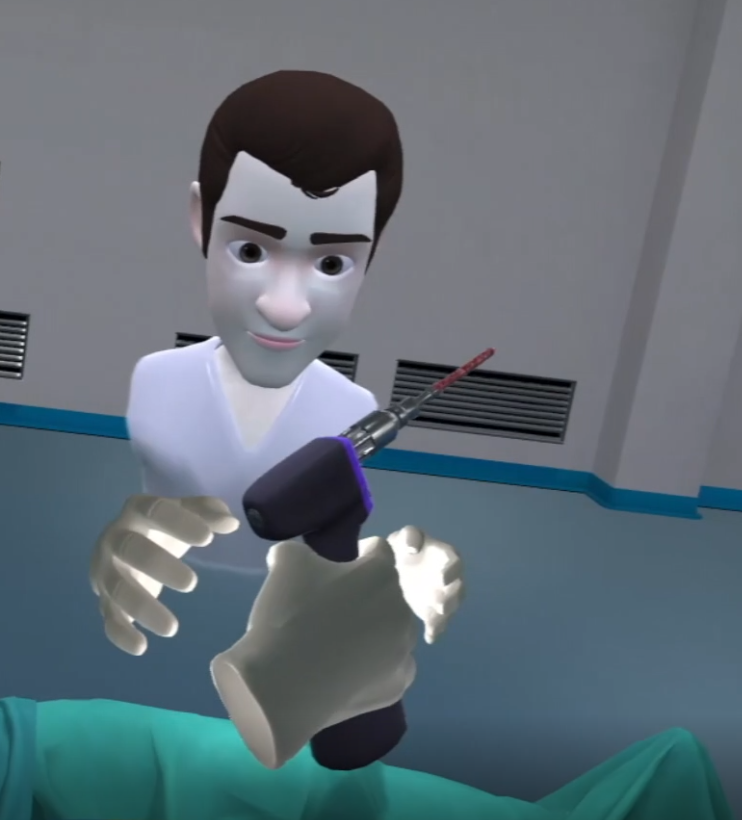}

  \caption{Images taken from a modern VR training application 
  that incorporates our proposed interpolation methods 
  for all rigid object transformations as well as hand 
  and avatar movements. It is recommended to see the video 
  presentation of this work \cite{kamarianakisCGI2021presentation}, to 
  better understand the significance of these figures.}
  \label{fig:flexing}
\end{figure}

\paragraph{Our Contribution.} In this work, we propose novel methods to 
address the two aforementioned problems; A)effective handling of data 
exchange in multi-user VR sessions and B) efficient recording of 
user actions in VR sessions, especially for training applications. 
For both problems, we deploy mathematical tools from 
geometric algebra (GA) and various (sub)algebras. 
Our methods rely on the fact that  all basic geometric 
primitives and their transformations used in VR, such as points, 
planes, lines, translations, 
rotations and dilations (uniform scalings), can be uniformly 
represented as \emph{multivectors}, i.e., elements of a suitable 
geometric algebra such as 3D Projective (3D PGA) or 
3D Conformal Geometric Algebra (3D CGA). 
Furthermore, we provide an alternative method that
utilizes dual quaternions as the 
representation form of the 
position and rotation of the users, both for transmission
and storing purposes.
Our two methods are compared in detail with the current state of 
the art (SoA), that deploys vectors and quaternions to manage 
positional and rotational data respectively. The comparison for both 
problems addressed is accomplished in networks of variable 
bandwidth capacity, rating from unrestricted networks to ones 
that are heavily limited. In all cases, we demonstrate that our 
methods perform equally or outperform SoA, with increased benefit 
from our methods as the network quality deteriorates (see Section~\ref{sec:our_results}). 
Regarding the data exchange problem, handled in 
Section~\ref{sec:transmission}, we provide convincing results 
in a modern game engine and a VR collaborative training 
scenario (see the video presentation of this work \cite{kamarianakisCGI2021presentation} and Figure~\ref{fig:flexing}). 
Our methods concerning VR recording are described in 
Section~\ref{sec:recording}, where we present evidence that 
using multivectors and (dual) quaternions as an alternative 
to the SoA representation forms yields similar results, especially
when data interpolation is required. 

\paragraph{Why use Geometric Algebra?}  
Algebras such as 3D PGA and 3D CGA\footnote{Of course, instead of 3D PGA-CGA one could also employ a, geometric 
or not, algebra that would allow the representation of motors in a form suitable for interpolation. Determining if such an algebra, non-isomorphic to a sub-algebra of 3D PGA or 3D CGA, exists was not done in the context of this work.} are showing rapid adaptation to VR
implementations due to their ability to represent the commonly used
vectors, quaternions and dual-quaternions natively as multivectors. 
In fact, quaternions and dual-quaternions are contained as 
a sub-algebra in both these algebras \cite{DietmarFoundations}. 
Therefore, they incorporate 
all benefits of quaternions and dual-quaternions representations such 
as artifact minimization in interpolated frames\cite{Kavan2008}. 
 Furthermore, 
geometric algebras enable powerful geometric predicates and modules
\cite{kamarianakis2021all}, providing, 
if used with caution, performance which is on par with the current state-of-the-art frameworks\cite{Papaefthymiou:2016dx}. In the past decades, 
these algebras have proven to be able to solve a variety of problems in
various fields, involving inverse kinematics \cite{hildenbrandInverseKinematicsComputation2008} and physics \cite{doran2003geometric}.
In conclusion, this work 
constitutes yet another step towards an effective all-in-one geometric 
algebra framework for handling VR data streams, with performance 
that is on par or exceeds current SoA frameworks.

\section{Basic Mathematical Notation}
In the following sections, we provide basic notation and formulas for 
(dual) quaternions, dual numbers and multivectors representing motors, for the sake of completeness. Similar or more complicated formulas can be obtained via \cite{Kavan2008,DietmarFoundations,DorstBook} or any other modern book involving dual quaternions and multivectors. We strongly encourage the reader to visit \url{https://bivector.net} for more information on these beautiful algebras.  

\subsection{Quaternions} 
\label{sub:quaternions}

The typical form of a quaternion $q$ is 
\begin{equation}
q:=q_w+q_x\pmb{i}+q_y\pmb{j}+q_z\pmb{k}, 
\end{equation}
where $q_w,q_x,q_y,q_z\in\mathbb{R}$ and the imaginary units 
$\pmb{i},\pmb{j},\pmb{k}$ satisfy the well known identities 
$\pmb{i}^2=\pmb{j}^2=\pmb{k}^2=\pmb{i}\pmb{j}\pmb{k}=-1$.
The quaternion $q$ can also be viewed as a 4-tuple $(q_w, q_x, q_y, q_z)$.
If $q_w$ equals $0$, the $q$ is called a \emph{pure quaternion}.

Addition of the quaternion $q$ to another quaternion 
\begin{equation}
p:=p_w+p_x\pmb{i}+p_y\pmb{j}+p_z\pmb{k}, 
\end{equation}
is done  componentwise, i.e., 
\begin{equation}
q+p:=(q_w+p_w)+(q_x+p_x)\pmb{i}+(q_y+p_y)\pmb{j}+(q_z+p_z)\pmb{k}.
\end{equation}
Multiplication of the two quaternions is carried out as follows: 
\begin{align}
qp:=&(q_wp_w-q_xp_x-q_yp_y-q_zp_z)+(q_wp_x+p_wq_x+q_yp_z-q_zp_y)\pmb{i}\\
&+
(q_wp_y+p_wq_y+q_zp_x-q_xp_z)\pmb{j}+(q_wp_z+p_wq_z+q_xp_y-q_xp_y)\pmb{k}.\nonumber
\end{align}
by applying the associative property, applying the basic identities and 
gathering terms. Note that multiplication is generally not commutative.

The conjugate of $q$ is the quaternion 
\begin{equation}
q^*:=q_w-q_x\pmb{i}-q_y\pmb{j}-q_z\pmb{k}, 
\end{equation}
and it is easy to verify that 
\begin{equation}
(pq)^* = q^*p^*.  
\end{equation}
The norm of $q$ is defined as 
\begin{equation}
|q| := \sqrt{qq^*}=\sqrt{q_w^2+q_x^2+q_y^2+q_z^2},
\end{equation}
and its inverse, assuming $|q|\neq 0$, is 
\begin{equation}
q^{-1}:= \frac{q^*}{|q|^2}.
\end{equation}

If $|q|=1$, $q$ is called a \emph{unit quaternion}, and it can be written 
in the form 
\begin{equation}
q = \cos{\frac{\theta}{2}} + \sin{\frac{\theta}{2}}(u_x\pmb{i}+u_y\pmb{j}+u_z\pmb{k}),
\end{equation}
where $\pmb{u}:=(u_x,u_y,u_z)$ is a unit vector and $\theta\in [0,\pi]$. 
Such quaternions \emph{encapsulate} the rotation of a 3D point
$(p_x,p_y,p_z)$ by angle $\theta$ around an axis going through the 
origin in the direction of $\pmb{u}$. 
Indeed, if we apply the sandwich quaternionic product to the pure 
quaternion $p=(0,p_x,p_y,p_z)$ we will obtain the pure quaternion
\begin{equation}
p' := qpq^* = (0,p'_x,p'_y,p'_z),
\end{equation}
where the $(p'_x,p'_y,p'_z)$ is the image of the point $(p_x,p_y,p_z)$
if we applied the aforementioned geometric transformation.


\subsection{Dual Numbers} 
\label{sub:dual_numbers}
A \emph{dual number} $d$ is defined to be 
\begin{equation}
d := a+\epsilon b,
\end{equation}
where $a,b$ are elements of some field (usually $\mathbb{R}$) and $\epsilon$ is a \emph{dual 
unit}, i.e., it holds that $\epsilon^2=0$. The elements $a$ and $b$ 
are referred to as the \emph{real} and \emph{dual} part of $d$. 

Let $d_i := a_i+\epsilon b_i$, for $i\in\{1,2\}$ be dual numbers. 
The addition of these dual numbers is performed pairwise, i.e., 
\begin{equation}
d_1+d_2 := (a_1+a_2) + \epsilon(b_1+b_2),
\end{equation}
and their multiplication is done by applying the associative rule and 
gathering terms
\begin{equation}
d_1d_2 := (a_1+\epsilon b_1)(a_2+\epsilon b_2) = a_1a_2+\epsilon (a_1b_2+a_2b_1),
\end{equation}
taking into consideration that $\epsilon^2=0$. Note that the multiplication on the right hand side of the equations above depend on the field where the
coefficients $a_1,b_1,a_2$ and $b_2$ belong.

The multiplication identity is $1+\epsilon0$ and the multiplicative inverse
of $d=a+\epsilon b$, when $a\neq 0$, is 
\begin{equation}
d^{-1} = a^{-1} (1-\epsilon ba^{-1}). 
\end{equation}
If $a=0$, then $d=\epsilon b$ has no inverse, which dictates that dual numbers form a ring and not a field. 

The conjugate of a dual number $d=a+\epsilon b$ is the dual number
\begin{equation}
\bar{d} = a - \epsilon b.
\end{equation}

A really interesting and useful property of dual numbers is that, the function of a dual number $d=a+\epsilon b$ can be obtained by considering the Taylor expansion at $a$, i.e.,
\begin{equation}\label{eq:f_dual_numbers}
f(a+\epsilon b) = f(a) + \epsilon b f'(a) + \frac{1}{2}\epsilon^2 b^2 f''(a)+\cdots = f(a) + \epsilon b f'(a).
\end{equation}


\subsection{Dual Quaternions} 
\label{sub:dual_quaternions}

A \emph{dual quaternion} is a dual number $D = p + \epsilon q$, where 
$p,q$ are quaternions. Regarding dual quaternions\footnote{As \emph{bi-quaternions} gain 
more momentum, we note here for clarity that they are distinct from 
dual-quaternions. Indeed,  dual quaternions are dual numbers where each 
coefficient is a quaternion whereas, the bi-quaternions are quaternions 
where each coefficient is a complex number. A basic difference is that,  
that apart from the common basic elements $\{1,i,j,k\}$, the additional 
element $I$ of bi-quaternions satisfies $I^2=-1$, whereas the additional 
element $\epsilon$ of the dual-quaternions satisfy $\epsilon^2=0$.}, the addition, multiplication and inverse are determined using the forms used in the 
dual numbers case, taking into consideration that $\epsilon$ commutes with 
the imaginary units $\pmb{i},\pmb{j}$ and $\pmb{k}$. 
If $p=(p_w,p_x,p_y,p_z)$ and $q=(q_w,q_x,q_y,q_z)$,
then $D$ can also be seen as the vector $(p_w,p_x,p_y,p_z,q_w,q_x,q_y,q_z)$.

There are two main conjugates of a dual quaternion $D = p + \epsilon q$ which are obtained via
\begin{itemize}
  \item the conjugation of the quaternions parts
  \begin{equation}
  D^* = p^* + \epsilon q^*,
  \end{equation}
  \item the conjugation of the dual number form
  \begin{equation}
  \bar{D} = p - \epsilon q.
  \end{equation}
\end{itemize}
We can easily verify that $\overline{D^*}=\left(\bar{D}\right)^*$. 

If $a,b$ are quaternions, we denote 
\begin{equation}
\langle a,b\rangle := \frac{1}{2}(ab^*+a^*b).
\end{equation}
The norm of $D = p + \epsilon q$ is then equal to
\begin{equation}
|D| := \sqrt{DD^*} = \sqrt{|p|^2+2 \epsilon \langle p,q\rangle} = 
|p|+ \epsilon \frac{\langle p,q\rangle}{|p|^2}.
\end{equation}
The last equality is obtained by applying ~\ref{eq:f_dual_numbers}, for 
$f(x)=\sqrt{x}$. If $D_1,D_2$ are dual quaternions, it holds that 
$|D_1D_2|=|D_1||D_2|$.

A dual quaternion $D = p + \epsilon q$ that satisfies $|D|=1$, or 
equivalently $|p|=1$ and $\langle p,q\rangle=0$, is 
called \emph{unit dual quaternion}. Due to these two linear restrictions, the set of unit dual quaternions, seen as vectors, form a 6-dimensional 
manifold embedded in the 8-dimensional Euclidean space. Also note that 
if $D_i$, for $i\in\{1,2\}$, are unit dual quaternions then $D_i^{-1}=D_i^*$ and the product $D_1D_2$ is also a unit dual quaternion.

A remarkable theorem (\cite[Lemma 12 in p12]{Kavan2008}) is the following. 

\begin{lemma}
Every rigid transformation can be represented by a unit dual quaternion, and conversely, every unit dual quaternion represents a rigid transformation.
\end{lemma}

The proof of the lemma is based on the fact that a unit dual quaternion
$D$ can be written in the form $D = p + \epsilon q$, where $p$ is a unit 
quaternion and
$q = \frac{1}{2}(t_1\pmb{i}+t_2\pmb{j}+t_3\pmb{k})$, for some vector 
$t:=(t_1,t_2,t_3)$. In this case, $D$ \emph{encapsulates} the rotation ``stored'' within $p$ followed by a translation by $t$. 
Indeed, if take the dual quaternion $v=(1,0,0,0,0,v_1,v_2,v_3)$ that 
\emph{corresponds} to a point $v$ and apply the sandwich multiplication by 
$D$, we will obtain 
\begin{equation}
v' = Dv\overline{D^*} = (1,0,0,0,0,v'_1,v'_2,v'_3)
\end{equation}
where $(v'_1,v'_2,v'_3)$ is the image of the point $(v_1,v_2,v_3)$ after applying the rotation encapsulated in $p$ and then translating by $t$.


\subsection{Motors in Geometric Algebra} 
\label{sub:motors_geometric_algebra}

In this section, we provide a brief overview of the 3D PGA or CGA
multivectors that arise in our methods. 
The multivectors studied represent motors, i.e., they are 
of the form $M=TR$, where $T$ and 
$R$ are the multivectors encapsulating a translation and a rotation,
respectively. Below, we demonstrate how we can obtain $T$ and $R$ from 
a given $M$, and how we can transmute them into the corresponding vector 
and unit quaternion form that correspond to the same geometric transformations. The later forms are needed by modern game engines to apply 
the transformation stored in $M$ to any game object. 

\begin{itemize}
\item {\textbf{3D PGA:}} In this algebra, the multivector
\begin{equation}
T := 1 -0.5e_0(t_1e_1+t_2e_2+t_3e_3),   
\end{equation}
represents the translation by $(t_1,t_2,t_3)$ and the multivector
\begin{equation}
R := a+be_{12}+ce_{13}+de_{23},
\end{equation}
encapsulates the same rotation with the unit quaternion 
\begin{equation}\label{eq:pga_multi_to_quat}
q:=a-d\pmb{i}+c\pmb{j}-b\pmb{k}. 
\end{equation}
In this algebra, it also holds that
\begin{equation}
e_0e_0=0,
\end{equation}
and therefore, if we evaluate the quantity $e_0M$,  we will obtain the multivector
\begin{equation}
e_0M\equiv e_0TR\equiv(e_0T)R\equiv e_0R.  
\end{equation} 
This multivector will be of the form
\begin{equation}
e_0R=ae_0+be_{012}+ce_{013}+de_{023},
\end{equation}
hence we can deduce that
\begin{equation}
R = a+be_{12}+ce_{13}+de_{23},
\end{equation}
which corresponds to the quaternion shown in \ref{eq:pga_multi_to_quat}.
The inverse of $R$ is the multivector
\begin{equation}\label{eq:R_inverse}
R^{-1} = a-be_{12}-ce_{13}-de_{23}.
\end{equation}

Since $M$ and $R$ are known, we may evaluate $T$, as it equals 
\begin{equation}
T=MR^{-1}=M(a-be_{12}-ce_{13}-de_{23})=1 +xe_{01}+ye_{02}+ze_{03},
\end{equation}
and conclude that it corresponds to a translation by $(-2x,-2y,-2z)$. 
\item{\textbf{3D CGA:}} In this algebra, the multivector
\begin{equation}
T=1-0.5(t_1e_1+t_2e_2+t_3e_3)(e_4+e_5) 
\end{equation}
is the multivector that corresponds to a translation by $(t_1,t_2,t_3)$.
Identical to 3D PGA, the multivector
\begin{equation}
R = a+be_{12}+ce_{13}+de_{23},
\end{equation}
encapsulates the same rotation with the unit quaternion 
\begin{equation}
q:=a-d\pmb{i}+c\pmb{j}-b\pmb{k}. 
\end{equation}
Since $M=TR$, it follows from the form of $T$ that
\begin{equation}
M=TR=R+m,
\end{equation}
where $m$ is a multivector that necessarily contains basis 
elements containing $e_4$ and $e_5$ (or their geometric product) 
that cannot be cancelled out. Therefore, we can obtain $R$ by 
keeping the terms of $M$ that contain \emph{only} the basis vectors  
$\{1,e_1,e_2,e_3,e_{12},e_{23},e_{13}\}$. 
Having determined $R$ and given $M$ is known, we 
evaluate $R^{-1}$ using \ref{eq:R_inverse} and determine $T$ using
\begin{equation}
T=MR^{-1}.
\end{equation}
Assuming $T$ is normalized (otherwise we normalize it), we may extract
the corresponding translation vector $(t_1,t_2,t_3)$ from the quantity 
$T(e_5-e_4)$, as it holds that
\begin{equation}
T(e_5-e_4)=t_1e_1+t_2e_2+t_3e_3.
\end{equation}
The conversion of $R$ to quaternion is identical with the case of 
3D PGA above.
\end{itemize}


\section{Data Transmission in Multi-User Collaborative VR Sessions } 
\label{sec:transmission}

In this section, we consider the transmission of a user's data to the 
rest of the users, in the context of a multi-user VR collaborative 
session. 
The information that is necessarily relayed over the network 
involves the users interactions 
through the hand-based HMD controllers such as \emph{displacement data} 
(e.g., translation and rotation of the controller) 
within specific time intervals and button-press
events. 
Since the VR rendering engine is typically different for each 
user, it is crucial to efficiently relay the data of one 
user to the rest of the users' VR engines, in a 
synchronized manner. 
If done correctly \cite{vilmi2020real}, all users will be
synchronized and actions between users, such as exchanging objects or 
even playing tennis, can be accurately performed. 
An inefficient transmission of this information will cause each 
user's virtual world to depict the rest of the users with increased
latency and therefore, collaboration will no longer be possible or 
feel natural.

To better understand the source of this problem, let us initially 
consider the pipeline of the data from a user to the rest.
When the user moves the hand-based
controllers of his HMD, the hardware 
initially detects the movement type and logs it, in various 
time intervals based on the user's or developer's preferences. 
This logged movement, that is either a translation and/or a rotation, 
is constantly transcoded into a suitable format and relayed 
to the VR application and rendered as a corresponding 
action, e.g., hand movement, object transformation or some action. 
The controller's data format to be 
transmitted to the rendering engine affects the overall 
performance and  quality of experience (QoE) and poses challenges 
that must be addressed.
These challenges involve keeping the latency between the movement 
of the controller and its respective visualization in the 
HMD below a certain threshold that will not break the user's  
immersion. Furthermore, the information must be 
relayed efficiently such that a continuous movement of the controller
results in a smooth jitter-less outcome in the VR environment. 
Such challenges heavily depend on the implementation details 
regarding the communication channel that handles the 
way that position and rotation of the controller is relayed, as well 
as the choice of a suitable interpolation technique. 
The displacement data are transmitted at discrete time
intervals, approximately 20-40 times per second. To maintain a high 
frame-per-second scenery in the VR, multiple in-between frames 
must be created on-the-fly by the appropriate tweening algorithm. An efficient 
algorithm will allow the generation of natural flow frames 
while requiring fewer intermediate keyframes. 
Such algorithms will help reduce a)
bandwidth usage between the HMD and the rendering engine and
b) CPU-strain, 
resulting in lower energy consumption as well as lower latency issues 
in bandwidth-restricted networks. Moreover, HMDs with 
controllers of limited frequency will still be able to deliver 
the same results as more expensive HMDs.


\subsection{State of the Art} 
\label{sec:state_of_the_art}


The current SoA methods regarding the format used to 
transmit the displacement data mainly involves the use of 3D vectors for 
translation and quaternions for rotation data. 
Modern game engines such as Unity3D and Unreal 
Engine, have these representation forms already built within 
their frameworks.
The dominance of these forms is based on the fact that 
they require very few floating point numbers 
to be represented (3 and 4 respectively) and their ability to support 
fast and efficient interpolations. Specifically, 3D vectors
are usually linearly interpolated, where as the SLERP method is 
usually used for quaternion blending. In some engines, such as 
Unity3D, rotations are sometimes provided in terms of Euler angles, but 
for interpolation needs, they are internally transformed to 
their quaternion equivalents.

Based on the current SoA forms, the controllers of a VR HMD 
log their current position $v=(v_x,v_y,v_z)$ at each time step with 
respect to a point they consider as the origin. 
Their rotation is also stored, as a 
unit quaternion $q=(q_w,q_x,q_y,q_z)=q_w+q_x\pmb{i}+q_y\pmb{j}+q_z\pmb{k}$. The use of unit quaternions 
revolutionized graphics as it 
provided a convenient, minimal way to represent rotations, 
while avoiding known problems (e.g., gimbal lock) 
of other representation forms such as Euler angles \cite{Kavan2008}. 
The ways to change between unit quaternions and 
other forms representing the 
same rotation, such as rotation matrices and Euler angles, 
are summarized in \cite{diebel2006representing}. 

The interpolation of the 3D vectors containing the positional data 
is done linearly, i.e., given $v$ and $w$ vectors we may generate 
the intermediate vectors $(1-a)v+aw$, for as many $a\in [0,1]$ as 
needed. Given the unit quaternions $q$ and $r$ the intermediate 
quaternions are evaluated using the SLERP blending, i.e., we evaluate
$q(q^{-1}r)^a$, for  as many $a\in [0,1]$ as 
needed, like before. If these intermediate quaternions are applied 
to a point $p$, the image of $p$, as $a$ goes from $0$ to $1$, 
has a uniform angular velocity around a fixed rotation axis, which 
results in a smooth rotation of objects and animated models.

\subsection{Room for Improvements} 
\label{sec:room_for_improvements}


  Graphics courses worldwide mention quaternions as the next evolution 
  step of Euler angles; a step that simplified things and 
  amended interpolation problems without adding too much overhead 
  in the process. Despite it being widespread, the combined use 
  of vectors and quaternions does not come without limitations. 

  A drawback that often arises lies in the fact that the 
  simultaneous linear interpolation of the vectors with the SLERP 
  interpolation of the quaternions applied to rigid objects does not 
  always yield smooth, natural looking results in VR. This is empirically 
  observed on various objects, depending on the movement the user 
  \emph{expects} to see when moving the controllers. Such  
  \emph{artifacts} usually require the developer's intervention 
  to be amended, usually by demanding more intermediate 
  displacements from the controller to be sent, i.e., by 
  introducing more non-interpolated keyframes. This results mainly 
  in the increase of bandwidth required as more information 
  must be sent back and forth between the rendering engine and the 
  input device, causing a hindrance in the networking layer. 
  Multiplayer VR applications, that heavily rely on the  
  input of multiple users on the same rendering engine for multiple 
  objects, are influenced even more, when such a need arises. 
  Furthermore, the problem is intensified 
  if the rendering application resides on a cloud or edge node; 
  such scenarios are becoming increasingly more common as they 
  are accelerated by the advancements of 5G networks and the 
  relative functionalities they provide. 


  \subsection{Proposed Method Based on Dual Quaternions} 
  \label{sub:proposed_method_based_on_dual_quaternions}

  In the past few years, graphics specialists have shown that 
  dual quaternions can be a viable alternative and improvement 
  over quaternions, as they allow us to unify the translation and 
  rotation data into a single entity. 
  Dual quaternions are created by quaternions if dual numbers 
  are used instead of real numbers as coefficients, i.e., they 
  are of the form $d:=A+\epsilon B$, where $A$ and $B$ 
  are ordinary quaternions and $\epsilon$ is the \emph{dual unit}, an 
  element that commutes with every element and satisfies 
  $\epsilon^2=0$ \cite{Kenwright:2012tl}. 
  A subset of these entities, called 
  \emph{unit dual quaternions}, are indeed isomorphic to the 
  transformation of a rigid body. A clear advantage of using 
  dual quaternions is the fact that we only need one framework 
  to maintain and that applying the encapsulated information 
  to a single point requires a simple sandwich operator. Moreover, 
  the rotation stored in the unit dual 
  quaternion $A+\epsilon B$ can be easily extracted as the  
  quaternion $r:=A$ is the unit quaternion 
  that amounts to the same rotation. Furthermore, if 
  $B^\star$ denotes the conjugate quaternion of $B$, 
  then $t:=2AB^\star$ is a pure quaternion whose coefficients 
  form the translation vector \cite{Kenwright:2012tl}. 

\begin{figure}[tbh]
  \centering
  \includegraphics[width=0.95\textwidth]{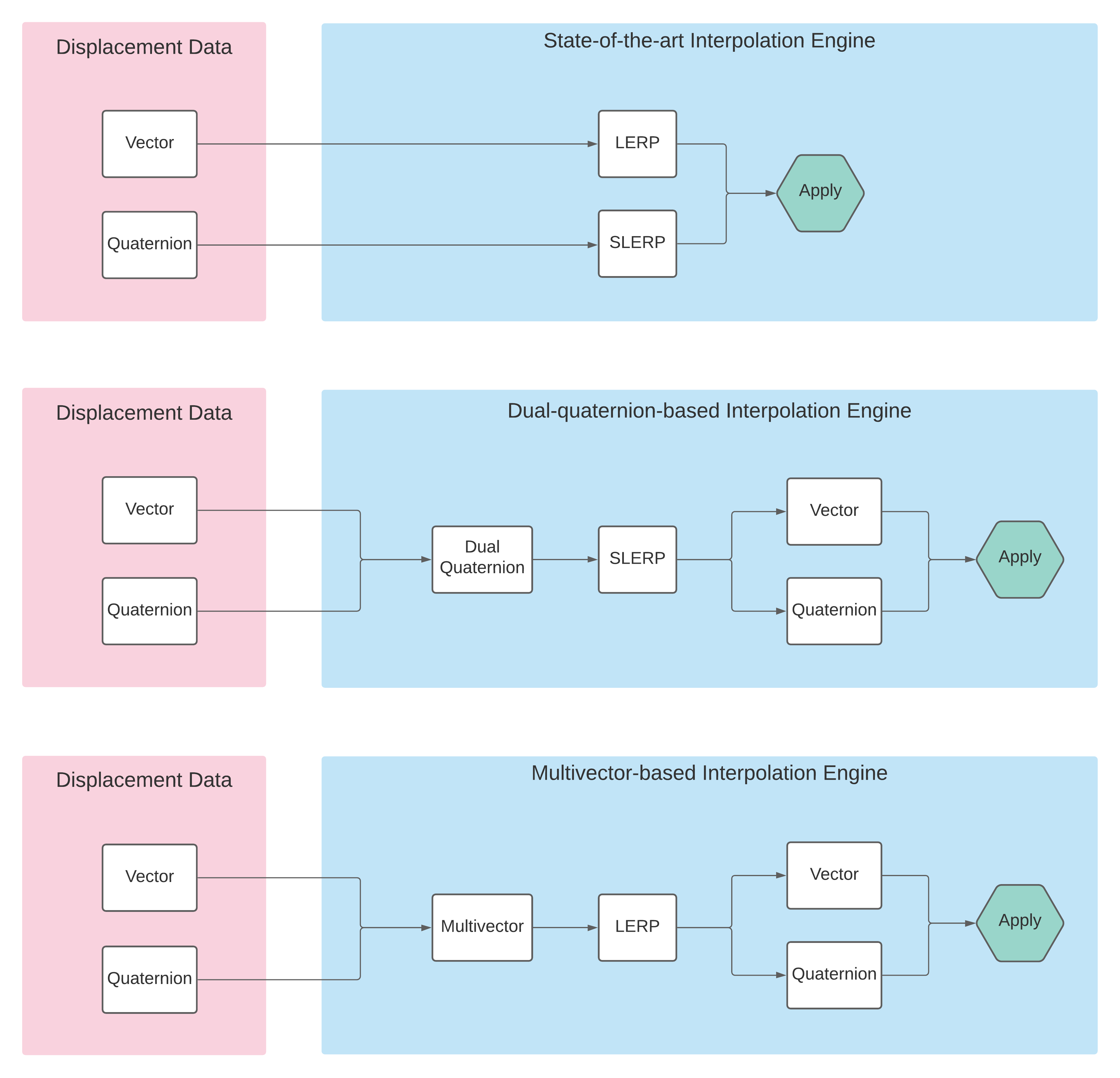}
  \caption{Algorithm layout of the different interpolation engines 
  used to generate intermediate frames.}
  \label{fig:sequence_diagrams}
  \end{figure}

  Taking advantage of the above, we propose the replacement 
  of the current state-of-the-art sequence 
  (see Figure~\ref{fig:sequence_diagrams}, Top) with 
  the following (see 
  Figure~\ref{fig:sequence_diagrams}, Middle). 
  The displacement data of an object is again represented as 
  a vector and a quaternion; in this way, only a total of 7 float 
  values (3 and 4 respectively) need to be transmitted. The VR engine 
  combines them in a dual quaternion \cite{Kenwright:2012tl}
  and interpolates with the previous state of the object, also 
  stored as a dual quaternion. Depending on the engine's and 
  the user's preferences, a number of in-between frames are 
  generated via SLERP interpolation \cite{Kavan2008} of    
  the original and final data. For each dual-quaternion received 
  or generated, we decompose it to a vector and a quaternion and 
  apply them to the object. This step is necessary to take advantage
  of the built-in optimized mechanisms and GPU implementations 
  of the VR engine. 

  A major advantage of the proposed method is that we can obtain 
  similar results with the state-of-the-art method by sending less 
  keyframes per second. As an empirical law, we may send $20$ 
  displacement data per second with our method to obtain the same 
  quality of generated frames as if we had sent $30$ data per second 
  with the current state-of-the-art method. This $33\%$ reduction 
  of required data applies for each user of the VR application, 
  greatly lowering the bandwidth required as more users join. 
  As an example, if $n$ users participate, the total displacement data 
  required for our method would be 560$n$ bytes per second
  ($20$ messages per second X $7$ floats per message 
  X $4$ bytes per double, assuming a classic implementation) as
  opposed to 840$n$ bytes per second ($30$ messages per 
  second X $7$ floats per message X $4$ bytes per float) 
  with the default method. 
  The numbers of updates per second mentioned above relate 
  to the case of unrestricted-bandwidth network; for the 
  respective results regarding constrained networks see Section~\ref{sec:our_results} and Table~\ref{tbl:results}.
  This method is incorporated in the MAGES SDK \cite{papagiannakis2020mages}
  for cooperative VR medical operations, publicly available to be tested for free, as the default transmission method, which is indicative of the 
  performance boost it provides.

  The drawbacks of this method is the need to constantly transform 
  dual-quaternions to vector and rotation data after every 
  interpolation step but this performance overhead is tolerable
  as the extraction of the displacement data is accomplished 
  in a straight-forward way. 
  Also, performing SLERP on a dual quaternion 
  (proposed method) instead of a quaternion (state-of-the-art method) 
  demands more operations per step. The trade-offs between 
  the two methods seem to favor our method, especially in the 
  case of collaborative VR applications.


  \subsection{Proposed Method Based on Multivectors} 
  \label{sub:proposed_method_based_on_multivectors}
  
\begin{figure}[t]
   \centering
   \includegraphics[width=0.95\textwidth]{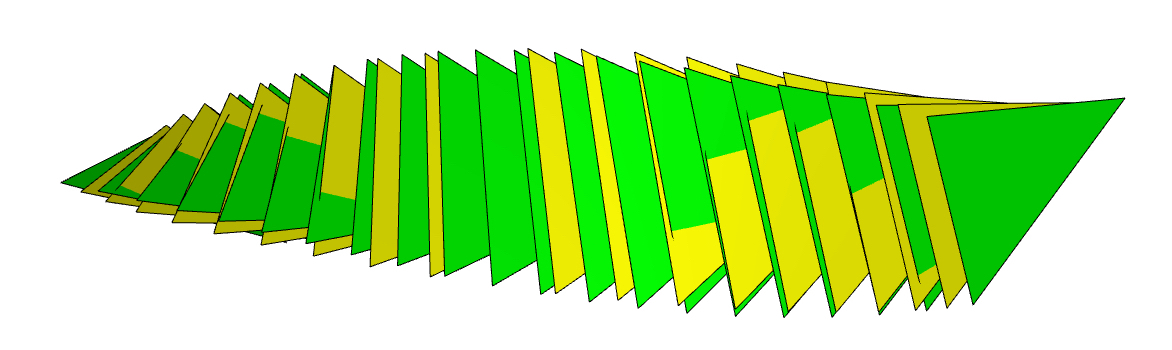}
   \caption{A triangular object is interpolated via multivectors. 
   A motor including both a translation and a rotation is applied 
   to the triangle via its mass center. Between the extreme positions
   of the object, 
   we generate 20 intermediate frames using LERP (yellow) and SLERP
   (green) interpolation of the multivector. Only minimal differences 
   are spotted between the two outcomes.}
   \label{fig:lerp_vs_slerp}
 \end{figure}

  The proposed method described in Section~\ref{sub:proposed_method_based_on_dual_quaternions} was based on the use 
  of dual quaternions and the fact that interpolating them (using SLERP)
  produced smooth intermediate frames. In this section, we go 
  one step further and suggest the use of multivectors instead of 
  dual-quaternions (see Figure~\ref{fig:sequence_diagrams},Bottom). 
  This transition can be done in a straight-forward way if we 
  use multivectors of 3D Conformal  
  (see \cite{DietmarFoundations}) or 3D Projective Algebra (see \cite{DorstBook} 
  and its updated Chapter 11 in  \cite{dorstguided}). The interpolation of the resulting 
  multivectors can be accomplished via LERP \cite{Hadfield:2019cx}; 
  if $M_1$ and $M_2$ 
  correspond to two consecutive displacement data, then we can 
  generate the in-between multivectors 
  \begin{equation}
  (1-a)M_1+aM_2,
  \end{equation}
  for as many $a\in [0,1]$ as needed (and normalize them if needed).
  Notice that since we are only 
  applying these displacements to rigid bodies, 
  we may use LERP instead of SLERP 
  (see Figure~\ref{fig:lerp_vs_slerp}).
  For every (normalized) multivector $M$ received or interpolated, 
  we may now 
  extract the translation vector and rotation quaternion, as shown in Section~\ref{sub:motors_geometric_algebra}. 
  Every multivector received 
  or generated has to be decomposed to a vector and a quaternion in 
  order to be applied to the object, as modern VR Engines natively 
  support only the latter two formats.

  The advantage of such a method lies on the fact that 
  we can use LERP blending of multivectors instead of SLERP. This 
  saves as a lot of time and CPU-strain; SLERP interpolation requires the 
  evaluation of a multivector's logarithm, which requires a lot 
  of complex operations \cite{Dorst:2011}. 
  Notice that, LERP is efficient in our case since
  only rigid objects displacements are transfered via the network; 
  if we wanted to animate skinned models via multivectors it is 
  known that only SLERP can produce jitter-less intermediate frames
  \cite{Kavan2008}. 
  Another gain of this proposed method is 
  the ability to incorporate it in an all-in-one GA framework, 
  that will use only multivectors to represent model, deformation and
  animation data. Such a framework is able to deliver 
  efficient results and embeds powerful modules 
  \cite{Papagiannakis:2013va,kamarianakis2021all,Papaefthymiou:2016dx}.
  In such frameworks, decomposition of multivectors to vectors and
  quaternions will be redundant, as we can apply the displacement 
  to the object's multivector form via a simple sandwich operation.

  The trade-offs of such an implementation are based on the fact 
  that modern VR engines do not natively support multivectors and 
  therefore production ready modules, with basic functions implemented,
  are almost non-existent.
  An exception is the Klein C++ module for 3D PGA, found in 
  \url{www.jeremyong.com/klein}; for 3D CGA no such module is 
  available the moment this paper is written. This makes it 
  difficult for GA non-experts to adopt and implement such methods. 
  Furthermore, multivectors require 
  16 (3D PGA) or 32 (3D CGA) float values to be represented and using 
  unoptimized, usually CPU and not GPU-based, modules to handle them 
  may result in slow rendering. 
  Optimized modules, such as GAALOP \cite{hildenbrand2010gaalop},
  can take advantage 
  of the fact that very few of the multivector coordinates are 
  indeed non-zero, as the multivectors involved are always  
  motors, i.e., represent translations and/or rotations, and therefore
  have specific form. 
  Since the full multivector algebra is not needed for this 
  application, various other approaches exist to achieve performance optimizations (e.g., exploiting the even and odd sub-algebras of motors), 
  even when starting from a formulation of the algorithm as a naive 
  full multivector formula.


\section{Recording and Replaying in VR}
\label{sec:recording}

 \begin{figure}[tbh]
   \centering
   \includegraphics[width=\textwidth]{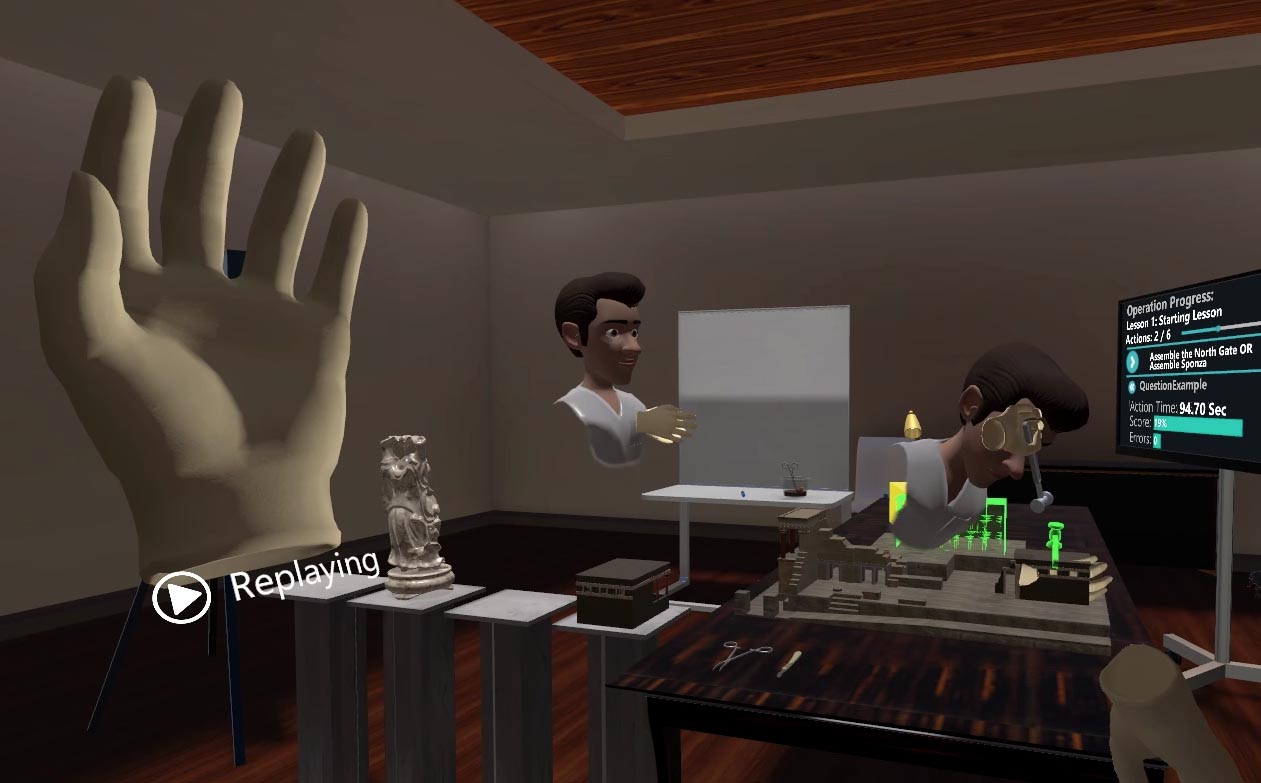}
   \caption{VR Record and Replay functionality. The point of view of the user replaying a recorded session. The user is able to watch the session from any perspective, as well as pause the replay in order to continue the session on his own.}
   \label{fig:replaying}
 \end{figure}

\subsection{Towards An Efficient VR Recorder} 
\label{sec:VR_Recorder}

Based on current bibliography, accurate recording of a VR session can be achieved via two methods, mainly differing in what they aim to log. 
The first method aims to record all users inputs whereas, the second method
focuses on the effects that these inputs have on the virtual scene.

As VR advances, the realism of the environments expands as well;
trivial interactions such as ripping a plastic case, grabbing a tool or 
removing the cap from a bottle, have turned to mandatory tasks that greatly 
add up to the immersive experience. 
This exponential content expansion is hindering the development of the 
latter logging method. 
Therefore, our research revolves around a VR logger that records raw users 
inputs, as we believe that such an approach suits better to the
current growth direction of virtual reality environments.

Opposed to popular recording techniques, we do not take into 
consideration video recording from cameras that are located in 
different places within the VR, whether these are static or, for 
various reasons, in motion.
Our decision is based on the fact that such recordings can not allow 
replaying the session from any other perspective.

Another major disadvantage of such recordings lies on the inability of the 
user to pause the recording and resume ``playing'' the session at a certain 
point. This is caused by the lack of storing values and data that are
deeply connected with the current state of the game at that time, but rather 
superficial snapshots of the VR scene. Without the ability 
of resume playing at a custom state, players cannot effectively improve 
their performance in a specific task, unless replaying the whole session. 

Raw video recording of the scene also introduces another major drawback. 
The position of objects and the trigger of various events from the VR 
is no longer easy to identify and keep track of, without extra effort. 
Even the use of signal processing algorithms and classification techniques, 
such as the ones described in \cite{Sharma2014, Zia2016, Zia2018}, will 
inevitably introduce errors and hinder the monitoring process of such data.

On the other hand, recording the  original input data of the users will
allow keeping track of the actual raw transformations of objects and events in the scene. This will sequentially enable methods for assessing the performance of users and understanding what the users are actually doing. 
Eventually, we should be able to create intelligent agents via imitation learning from experts that are able to complete a session ``successfully'', 
and support the users to effectively perform similar tasks accordingly.

Moreover, by avoiding collecting data of high dimensionality, such as the 
entire state of the virtual world and instead storing low-dimensional information such as the users inputs and triggers, we can 
obtain valuable analytics even by using simple processing algorithms. 
One should also notice that the data we choose to record can easily generate the scene by applying the worlds mechanics, while 
the reverse process would demand sophisticated computer vision algorithms. 
Lastly, using our recording method, users have the ability to 
act simultaneously with various recorded interactions and events, 
a functionality that can be used creatively to increase the pedagogical 
benefits of various simulations.

\subsection{Implementation Details} 
\label{sub:implementation_details}

In our proposed recording method, the displacement of all users in the operation is captured, along with their interactions with virtual objects.
The voice of each player is recorded individually, including 
incoming voice in cases of multi-player sessions.
Additionally, the actions performed as well as the complete 
traversal path of the scenegraph 
\cite{zikasSceniorImmersiveVisual2020} are captured.
In VR, the objects which the user interacts with might come into 
contact or interact with other objects, changing the latter's 
location or status. 
Therefore, the transforms of all subsequently affected objects 
are also recorded.

Four scripts are responsible for recording all the necessary data 
mentioned above; the \textsf{InteractionRecorder},
the \textsf{PropagateRecording}, the \textsf{GetAudioSamples} and 
the \textsf{RecordingWriter} scripts. 
The \textsf{InteractionRecorder} tracks each user's head 
and hands transforms, while the \textsf{PropagateRecording} 
records the status changes of every object that 
users interact with. 
All sound related recordings are received by the \textsf{GetAudioSamples}
script.
Finally, the \textsf{RecordingWriter} is responsible for writing all 
recorded interactions, sounds and events from previous three scripts 
into respective files. Note that an instance of each script 
is created for every user of a multi-player session.

We denote the files used to store the transforms of the 
head, hands and moved objects by \textsf{Transform \emph{X}}, whereas
the files that store all the events (\emph{messages}) 
occurring in the session 
are denoted by \textsf{Messages \emph{X}};
in both cases, \textsf{\emph{X}} is either \textsf{Camera}, 
\textsf{Left Hand}, or \textsf{Right Hand}.
Finally, the \textsf{RecordingInfo} file 
stores information about the recording, such as the duration of 
the recording, and is available only for the \emph{owner} of the 
session, i.e., the (first) user that initiated the session.

 \begin{figure}[tbh]
   \centering
   \includegraphics[width=0.8\textwidth]{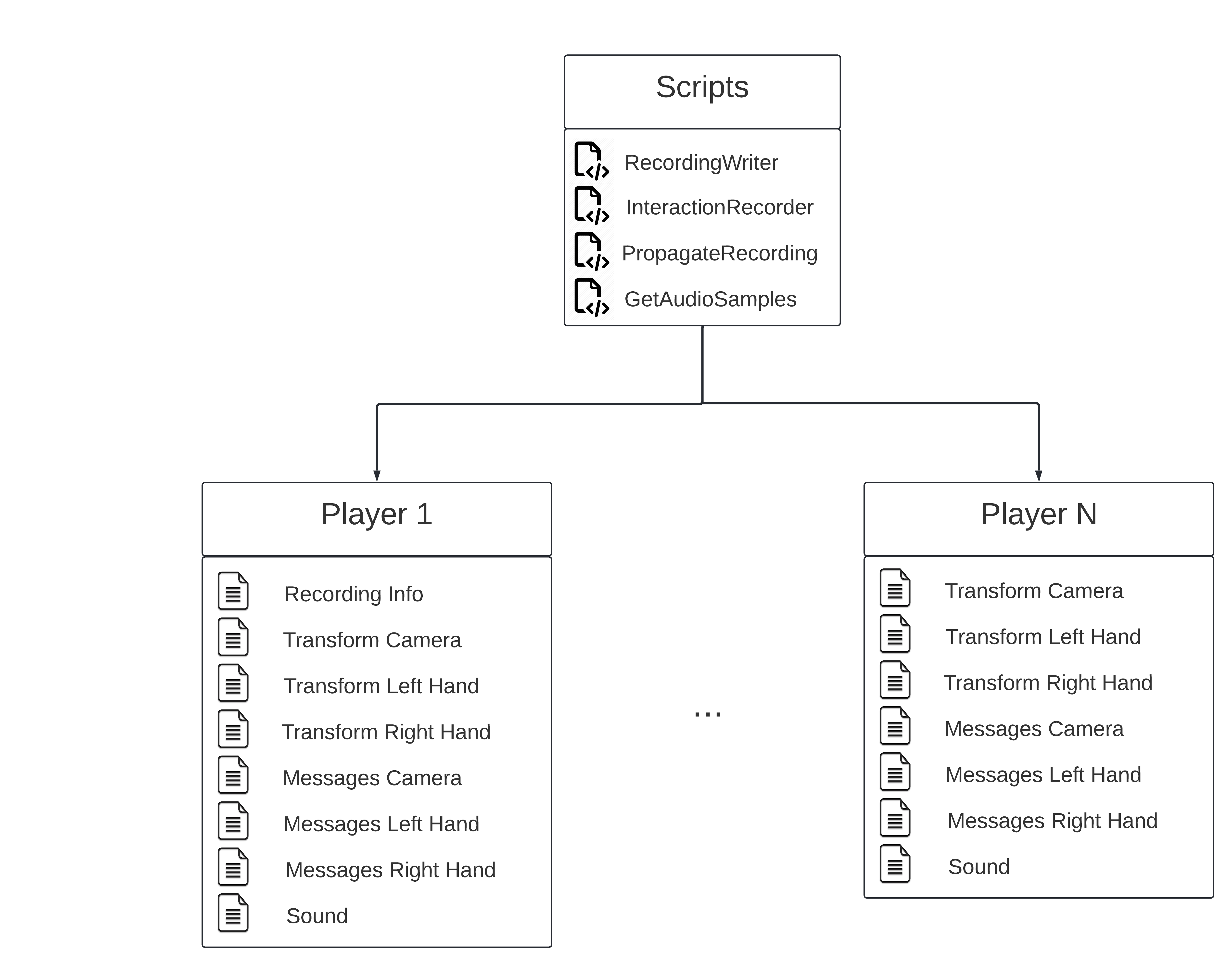}
   \caption{The associated scripts along with the generated files,  
   in the context of recording a VR session of N players.}
   \label{fig:VR_Recorder}
 \end{figure}

Below we present a list of all the data that our proposed VR Recorder 
stores. The data types used are either 3D Vectors (for transformation data), 
Events (for various Scene events and functionalities) or float (for the 
time).

\begin{itemize}
\item \textbf{Left/Right Hand Translation (3D Vectors):} Consists of 6 floats (3 for each hand) describing the position of each hand in the virtual environment
\item \textbf{Left/Right Hand Rotation (3D Vectors)}: Consists of 6 floats (3 for each hand) describing the rotation of each hand in the virtual environment
\item \textbf{Camera Translation (3D Vector):} Consists of 3 floats describing the position of the user’s HMD in the VR environment
\item \textbf{Camera Rotation (3D Vector):} Consists of 3 floats describing the rotation of the user’s head in the VR environment
\item \textbf{Left/Right Hand Start Interaction (Event):} The user has started interacting with an object. Consists of the name of the interacted object (string).
\item \textbf{Left/Right Hand End Interaction (Event):} The user has ended interacting with an object. Consists of the name of the interacted object (string), and the amount of time the user was interacting with the object (float).
\item \textbf{3D Objects Translation and Rotation (3D Vectors):} Consists of 6 floats describing the position and rotation of an object, with which a hand has interacted.
\item \textbf{Press Button (Event):} Some 3D objects provide an extra functionality while holding them, such as pressing the button for activating a drill. 
\item \textbf{Scenegraph Traverse (Event):} Information about traversing the scenegraph (e.g. go to next/previous action/stage/lesson). 
\item \textbf{Time (float):} The time that has passed since the start of the session, in seconds.
\end{itemize}

Figure~\ref{fig:VR_Recorder} depicts the
VR Recorder pipeline, showing the different scripts 
that are involved as well as the files that are created.
During replay, all ``replay'' avatars and interacted objects 
are recreated and referenced based on these recorded files. 
The replay functionality is handled by the \textsf{Replay} script.

\subsection{The Basic Algorithms}
\label{sub:basic_script_algorithms}

In this section, the algorithms behind the four basic scripts involved in  VRRR are presented. 

\textsf{RecordingWriter} (see Algorithm~\ref{alg:recording_writer_algo}) is responsible for writing all information regarding the position and the rotation of the logged objects, as well as messages of events relevant to them. 
The \textsf{RecordingWriter} is implemented using the singleton pattern, i.e., only one script exists on each scene.
Algorithm~\ref{alg:recording_writer_algo} provides the implementation of the \textsf{Recording Writer} script. 
In this algorithm, the \textsf{InteractionRecorder} script is added in the head and hands of all players when recording is started. 
The tracking objects, i.e., heads and hands, have two corresponding files where their transformations and messages relevant to them are stored. 
The \textsf{WriteToFile} public function is called by the \textsf{InteractionRecorder} and \textsf{PropagateRecording} scripts in order to write the information mentioned above. 
Finally, the \textsf{EndRecording} function closes all files when the session is finished.

\begin{algorithm}[tbh]
    \caption{Recording Writer}
    \label{alg:recording_writer_algo}
    \begin{algorithmic}[1]
        \State \textbf{Start():}
        \ForEach{player}
        \State Add Interaction Recorder Script in head and hands (tracking objects)
        \ForEach{tracking object}
            \State Create files  for storing transforms and messages
        \EndFor
        \EndFor
        \newline
        \State \textbf{WriteToFile(}tracking object, content, player\textbf{):}
        \State Check whether content contains a message or a transform
        \If{content contains message} 
            \State Write content in messages file for the tracking object of player
        \ElsIf{content contains transform} 
            \State Write content in transforms file for the tracking object of player
        \EndIf
        \newline
        \State \textbf{EndRecording():}
        \State Close Transform and Message Files for each player
    \end{algorithmic}
\end{algorithm}

\textsf{InteractionRecorder} (see Algorithm~\ref{alg:interaction_recorder}), is responsible for capturing the transformations on each frame of the tracking object that this script is attached to, as well as all the events that are relevant.
The \textsf{RecordingWriter} passes the ID of the player this game object is attached to, in order to send the corresponding information to the associated file.
Afterwards, the type of the tracking object is found and, according to that, specific listeners are initialized. 
These listeners are responsible for sending messages to the \textsf{RecordingWriter} when specific events have been performed. 
Such kinds of events are interactions of hands with other objects or tools, or changes in the state of the scene, which are stored in the head's messages file. 
The \textsf{Update} function is responsible for sending the transformation of the tracking object to the \textsf{RecordingWriter}, and, also, check if an event has occured in order to write it in the relevant file.
Finally, the \textsf{OnBeginInteraction} and \textsf{OnEndInteraction} functions add or remove the \textsf{PropagateRecording} script to the game object the tracking object has started or finished interacting with, and, also, send a relevant message to the \textsf{RecordingWriter}.

\begin{algorithm}[tbh]
    \caption{Interaction Recorder}
    \label{alg:interaction_recorder}
    \begin{algorithmic}[1]
        \State Player $\gets$ the ID of the player this gameobject refers to 
        \newline
        \State \textbf{Start():}
        \If{this gameobject is Hand} 
            \State Tracking Object $\gets$ Hand
        \ElsIf{this gameobject is Head} 
            \State Tracking Object $\gets$ Head
        \EndIf
        \If{Tracking Object is Hand} 
            \State Add listener OnBeginInteraction()
            \State Add listener OnEndInteraction()
        \ElsIf{Tracking Object is Head} 
            \State Initialize Event Messages Functions
        \EndIf
        \newline
        \State \textbf{Update():}
        \State Content $\gets$ Transform of this gameobject
        \State RecordingWriter.WriteToFile(Tracking Object, Content, Player)
        \If{Event has been executed} 
            \State Content $\gets$ Event Message
            \State RecordingWriter.WriteToFile(Tracking Object, Content, Player)
        \EndIf
        \newline
        \State \textbf{OnBeginInteraction(}Interacted GameObject\textbf{):}
        \State Add PropagateRecording script on Interacted Gameobject
        \State Content $\gets$ Start Interaction Message
        \State RecordingWriter.WriteToFile(Tracking Object, Content, Player)
        \newline
        \State \textbf{OnEndInteraction(}Interacted GameObject\textbf{):}
        \State Remove PropagateRecording script from Interacted GameObject
        \State Content $\gets$ End Interaction Message
        \State RecordingWriter.WriteToFile(Tracking Object, Content, Player)
    \end{algorithmic}
\end{algorithm}

\textsf{PropagateRecording} (see Algorithm~\ref{alg:propagate_recording}) sends the transformations of all the game objects that the hands have interacted with to the \textsf{RecordingWriter}. 
At first, information about the player and the game object is stored. 
Then, a listener is added to that object, if it is a tool. 
This listener is responsible for adding the \textsf{PropagateRecording} script to the objects the tool has interacted with. 
Finally the \textsf{Update} function destroys the script, if the object has stopped moving, and, also, writes the transformation data using the player's ID and the object's name as parameters in order for the \textsf{RecordingWriter} to find the associated file. 

\begin{algorithm}[tbh]
    \caption{Propagate Recording}
    \label{alg:propagate_recording}
    \begin{algorithmic}[1]
        \State Player $\gets$ the ID of the player this gameobject refers to
        \State Tracking Object $\gets$ this Gameobject
        \newline
        \State \textbf{Start():}
        \If{Gameobject is Tool} 
            \State Add listener OnToolBeginInteraction on Interacted Gameobject
        \EndIf
        \newline
        \State \textbf{OnToolBeginInteraction(}Interacted Gameobject\textbf{):}
        \State Add PropagateRecording script on Interacted Gameobject
        \newline
        \State \textbf{Update():}
        \If{Gameobject is not moving} 
            \State Destroy this script
        \EndIf
        \State Content $\gets$ Transform of this Gameobject
        \State RecordingWriter.WriteToFile(Tracking Object, Content, Player)
    \end{algorithmic}
\end{algorithm}

Finally, \textsf{Replay} (see Algorithm~\ref{alg:replay}) is utilized when a user chooses to watch a replay of a recorded session. 
This script is also a singleton, and all the other scripts access it by referring to it's instance.

At start, all event actions are coupled with the associated messages which are either predefined, or have been added by the developer for a specific application. 
Thus, when \textsf{Replay} reads a message in the messages files, it knows which event to execute. 
For example, when the \textsf{Start Interaction} message is read, the algorithm calls the event that is triggered when an interaction of a hand with a game object has begun. 
This event executes the same commands that are performed when users are in play mode. 
Therefore, the effect of the message is the same with having the avatar of a user perform a specific action.
After the initialization of the event actions, the algorithm reads the information of the recording, such as if it is single-player or multiplayer. 
Afterwards, the wait time of each player is calculated, in order to synchronize all the players' avatars, because the users may have entered the session in different times. 
Also, this wait time is used for synchronizing the graphics with the sounds captured from the players' microphones. 
At last, the \textsf{Start} function opens all the transformation and messages files, in order to read their content on each frame on the \textsf{Update} function.

On each frame, the following commands are executed for each player. 
Firstly, waiting or skipping the graphics update is executed. 
This wait or skip ensures that the graphics are synchronized with the sounds of each player's avatar.
After that, the script reads the next line of the transformation and messages files and check if the sound and graphics are de-synchronized.
Finally, a check whether an event has to be executed is performed.
If there is an event for this frame, the message is executed. 
Otherwise, the transformation changes are applied to the hands, heads and game objects.

\begin{algorithm}[tbh]
    \caption{Replay}
    \label{alg:replay}
    \begin{algorithmic}[1]
        \State \textbf{Start():}
        \State Initialize all Event Actions
        \State Get recording information
        \ForEach{Player} 
            \If {Session is Multi-Player} 
                \State Get Player's wait time
            \EndIf
            \State Delay $\gets$ Player's wait time
            \State Start Player's Sound File with Delay
            \State Open Player's Transformation and Messages files
        \EndFor
        \newline
        \State \textbf{Update():}
        \ForEach{Player} 
            \If{Player Graphics have to wait} 
                \State Wait graphics for Player
            \EndIf
            \If{Player Graphics have to move forward} 
                \State Skip graphics updates for Player
            \EndIf
            \State Read Transform Files
            \State Check if Sound and Graphics are Desynchronized
            \State Read Message Files
            \If{Event has to be executed} 
                \State Execute Message
            \ElsIf{No Event to Execute} 
                \State Apply Transforms on Hands, Head and Gameobjects
            \EndIf
        \EndFor
    \end{algorithmic}
\end{algorithm}

\subsection{VR Recording and Replay as part of a modern game engine} 
\label{sub:mages}

The VR Recording and Replay (VRRR) functionality introduced in this work is 
already implemented in the MAGES SDK 
\cite{papagiannakis2020mages,mages2018}, developed by ORamaVR and 
available for public testing on ORamaVR's website \cite{oramavr}. 
A poster version of this work was recently accepted in SIGGRAPH \cite{kamarianakis2022poster}; a video summarizing this work can be found in \url{https://youtu.be/_aoEAOzlyPg}.
MAGES is a novel VR authoring SDK,
scoping in accelerating the creation process of surgical training 
and the assessment of virtual scenarios. 
It is built on top of both the Unity3D and Unreal game engines and is composed of the following basic pillars:
\begin{itemize}
\item Multiplayer: collaborative networking layer that utilizes GA interpolation for bandwidth optimization.
\item Assessment: real-time performer assessment both with supervised machine learning and predefined rule-based analytics.
\item Deformations: GA-enabled deformable cutting and tearing, as well as configurable soft body simulations.
\item Curriculum: Tools for defining an educational curriculum enriched with visual guidance, gamified elements and objectives to enhance transfer of knowledge and skills.
\item Prototyped surgical techniques: Implementation of commonly used surgical techniques that can be customized in order to populate new content in a rapid manner.
\end{itemize}
The VRRR functionality extends the second 
and fourth pillars of the MAGES SDK, by enabling a) experts to record 
their sessions, b) novices to learn how to correctly perform an operation 
by watching the expert's recording and reviewing their own sessions, 
and c) evaluators to assess the learning outcomes of the apprentices 
by evaluating the users via the use of VR Replay 
(see Figure~\ref{fig:replaying}).

Via VRRR, we may record and replay a VR medical operation, in both single player and multi player modes. 
These recordings can be synchronized with the cloud and also be replayed on any device regardless of the original hardware they were recorded on.
This functionality is not just a video recording of the in-game view, but rather a full recreation of the operation as it happened when it was recorded. 
When replaying, the users are free to move around the operation room and watch from any angle they like.
The VR Recorder allows to log the user's sessions within a virtual environment in the form of positions, rotations and interactions, resulting in improved accuracy without compromising generalization. 
Additionally, the accurate data recorded enable the playback feature. 
This feature can be a applied for creating high fidelity VR replays that guide the trainee through his/her tasks. 

\section{Our Results} 
\label{sec:our_results}

\subsection{Metrics Regarding Transmission} 
\label{sub:metrics_regarding_transmission}

The methods proposed in Section~\ref{sec:transmission}
were implemented in Unity3D and applied to 
a VR collaborative training scenario. Figure~\ref{fig:catching} 
illustrates an example of extreme hand-based interpolation in collaborative, 
networked virtual environments. In order to properly understand the significance of this 
figure, it is advised to  watch the paper's presentation found in \cite{kamarianakisCGI2021presentation},
where we  demonstrate the effectiveness of our methods compared 
to the current state of the art. Specifically, we compare 
the three methods under different input rates per second, i.e., 
the keyframes sent per second to the VR rendering engine. 
The input rates tested are 5,10,15 and 20 frames per second (fps),
where the last option is an optimal value 
to avoid CPU/GPU strain in collaborative VR scenarios. 
These rates are indicative values of the maximum possible fps 
that would be sent in a network whose bandwidth rates from 
very-limited (5 fps) to unrestricted (more than 20 fps). In lower fps, 
our methods yield jitter-less interpolated frames compared to 
the state-of-the-art method, which would require 30 fps to 
replicate similar output. As mentioned before, this 
reduction of required data that must be transfered per second 
by 33\%-58\% (depending on the network quality, see 
Table~\ref{tbl:results}) is multiplied by every active user, 
increasing the impact and the effectiveness of our methods in bandwidth-restricted environments.

\begin{figure}[tbh]
   \centering
   \includegraphics[width=0.7\textwidth]{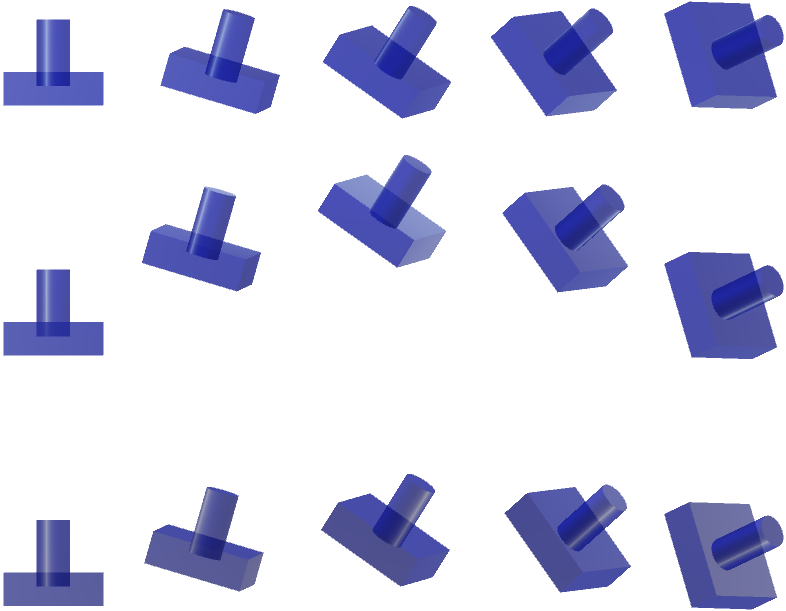}
   \caption{Different interpolation algorithms yield different, 
   yet jitter-less, intermediate frames. 
   (Top): State of the art: Vector and quaternion separate interpolation.
   (Middle): Dual-quaternion based interpolation algorithm.
   (Bottom): Multivector based interpolation algorithm.}
   \label{fig:interpolation}
 \end{figure}

The workflows of the two 
methods, compared with the current state of the art, are 
summarized in Figure~\ref{fig:sequence_diagrams}.
In Figure~\ref{fig:interpolation} we demonstrate the interpolation 
of the same object, at specific time intervals, for all methods; 
the intermediate frames feel natural for both methods proposed. 
\begin{table}[tbh]
  \caption{Summary of the metrics of our methods (Ours) versus 
  the state-of-the-art methods (SoA). The first  column 
  describes the possible network quality which correlates 
  to the maximum number of updates per second that can be performed. 
  The second column contains the update rate required to 
  obtain the same QoE under the specific network quality limitations.
  The third column contains the comparison of the bandwidth  
  and the running time difference by our algorithms compared 
  with the SoA algorithm, when using the respective update 
  rates of the second column.}
  \begin{center}
  \begin{tabular}{|c|c|c|}
  \hline
  Network Quality & How to Achieve Best QoE & Metrics on Our Methods \\
  \hline
  \hline
  \multirow{2}{*}{Excellent} & SoA: 30 updates/sec & 33\% less bandwidth\\
  & Ours: 20 updates/sec & 16.5\% lower running time\\
  \hline 
  \multirow{2}{*}{Good} & SoA: 20 updates/sec & 50\% less bandwidth\\
  & Ours: 10 updates/sec & 16.5\% lower running time\\
  \hline 
  \multirow{2}{*}{Mediocre} & SoA: 15 updates/sec & 53\% less bandwidth\\
  & Ours: 7 updates/sec & 16.5\% lower running time\\
  \hline 
  \multirow{2}{*}{Poor} & SoA: 12 updates/sec & 58\% less bandwidth\\
  & Ours: 5 updates/sec & 16.5\% lower running time\\
  \hline 
  \end{tabular}
  \label{tbl:results}
  \end{center}
\end{table}

In Table~\ref{tbl:results}, it is demonstrated that, 
under various network restrictions, both
proposed methods required less data (in terms of updates per sec) 
to be transmitted via the network to achieve the same QoE.
This decrease in data transfer leads to a lower energy 
consumption of the HMDs by 10\% (on average, preliminary result) 
and therefore enhances 
the overall mobility of the devices relying on batteries. 
To measure this increase in battery life, we used various 
untethered VR HMDs (HTC Vive Focus 3, Meta Quest 2,  Pico neo 3) 
where we run similar VR sessions with multiple (20+) users. 
In each session, we changed the transmission method and the updates 
interval rate as shown in Table~\ref{tbl:results}, in order to maintain 
the same QoE. 
For each network quality level, we noted and compared the battery 
drain difference of the same HMD, after a fixed period of time, between the 
SoA method and our methods with the respective update setting. For example, 
in the ``Excellent'' tier, we compared the battery drain, after 20 minutes, 
of the same HMDs, running the SoA method with 30 updates/sec against HMDs running the proposed methods with 20 updates/sec.

Our methods provide a performance boost, decrease the required time
to perform the same operation, with fewer keyframes but the same 
number of total generated frames, by 16.5\% 
(on average). The running times were produced in a PC with
a 3,1 GHz 16-Core Intel Core i9 processor, with 32 GBs of DDR4 memory. 
The same percentage of performance boost is expected in 
less powerful CPUs; in this case, the overall impact, in terms of 
absolute running time, will be even more significant.

\subsection{Metrics Regarding VR Recording} 
\label{sub:results_recording}

An examination of the performance of the VR Recorder was 
performed, and the impact on the computed frames per second 
(FPS) was assessed. 
For the evaluation, the frame rate (minimum, maximum and average) 
in two simulations was recorded, 
one with the VR Recorder disabled and one with the functionality enabled; 
the results are reported in table \ref{tbl:recorder_performance}. 
In order to provide accurate results, similar events were triggered. 
From the obtained results, one can discern that only a minor, 
unnoticeable by the human eye, drop in FPS occurred by enabling 
the VR Recorder, proof that the overall user experience was not 
affected negatively.

\begin{table}[tbh]
  \caption{Mean relative errors of interpolated translations for 1-frame and 2-frame interpolation, using Euclidean norm.}
  \begin{center}
  \begin{tabular}{|c|c|c|}
  \hline
  Algebra & 1-frame error (\%) & 2-frame error(\%)\\
  \hline
  \hline
  Dual Quaternions & 0.02 & 0.009\\
  \hline
  Linear Algebra & 0.006 & 0.009\\
  \hline
  3D CGA & 0.006 & 0.009\\
  \hline 
  \end{tabular}
  \label{tbl:recorder_intep_trans}
  \end{center}
\end{table}

 \begin{table}[tbh]
  \caption{Mean relative errors of interpolated rotations for 1-frame and 2-frame interpolation, using Euclidean norm.}
  \begin{center}
  \begin{tabular}{|c|c|c|}
  \hline
  Algebra & 1-frame error (\%) & 2-frame error(\%)\\
  \hline
  \hline
  Dual Quaternions & 0.02 & 0.02\\
  \hline
  Linear Algebra & 0.01 & 0.02\\
  \hline
  3D CGA & 0.01 & 0.02\\
  \hline 
  \end{tabular}
  \label{tbl:recorder_intep_rot}
  \end{center}
\end{table}

\begin{table}
  \caption{Measuring the FPS burden of a VR application due to 
  the Recorder feature. Evaluation of the average, minimum 
  and maximum FPS for a simulation with the functionality disabled 
  (Column 2) or enabled (Column 3).These results were obtained 
  by running simulations in a PC with an i7-11375H CPU, 16 GB of 
  RAM and an RTX 3060 GPU.}
  \begin{center}
  \begin{tabular}{|c|c|c|c|}
  \hline
  \multirow{2}{*}{Metric}  & Session without & Session with & \multirow{2}{*}{Difference} \\
  & VR Recording & VR Recording &  \\
  \hline
  \hline
  Average FPS & 89.56  & 85.13  & 4.43 \\
  \hline
  Minimum FPS& 76.56  & 68.78  & 7.78 \\
  \hline
  Maximum FPS& 93.29  & 92.57  & 0.72 \\
  \hline 
  \end{tabular}
  \label{tbl:recorder_performance}
  \end{center}
\end{table}

Since FPS is affected by constantly storing the translational and 
rotational data to files, we can further reduce the FPS load caused by 
the VR recorder by ``skipping'' data of several frames and instead 
generate them via interpolation, if later required. 
Such an approach will inevitably lead to severe reduction of the 
size of the stored data, depending on how many we skip. 
For example, if we choose to not record data on every other frame, 
we will have $50\%$ lower size, whereas if we choose to keep data 
once per three frames we will need only $1/3$ of the original storage 
space. 

Of course, as we record less frames, we are susceptible to 
interpolation problems and artifacts may arise. Dealing with 
the interpolation of the stored data, in case of replaying,
can be accomplished in the same way we dealt with
the interpolation of the data transmitted over the network, in 
Section~\ref{sec:transmission}. As proven in that section, 
different representation forms for the same, stored, displacement data 
require different pipelines to be interpolated (see Figure~\ref{fig:sequence_diagrams}). Apparently, the results regarding the quality 
of the interpolated data depending on their form, presented in 
Section~\ref{sec:our_results} and \cite{kamarianakisCGI2021presentation} also apply for the stored data.

To further demonstrate that using multivectors and/or dual quaternions
to store positional and rotational data instead of the classic 
SoA forms, we performed two additional experiments. In both experiments, 
we recorded a VR session and stored the data both in SoA form, i.e, vectors for translation and Euler angles for rotation, as well as in 
multivector and dual-quaternion forms. In the first experiment (A), we 
deleted the data of every other frame whereas, in the second experiment (B), 
we kept the data of one every three frames. In this way, we emulated 
the data that we would obtain if indeed we recorded one every two or three
frames respectively.

Depending on the representation form, we used the methods described in 
Sections~\ref{sub:proposed_method_based_on_multivectors} and 
\ref{sub:proposed_method_based_on_dual_quaternions} to generate the 
``missing'' data via interpolation ($a=1/2$ for experiment A and $a=1/3$ 
and $2/3$ for experiment B). The data created in this way were 
then compared to the originally recorded data by performing 
the following steps. For every pair of data that we wanted to compare, 
we extracted the encapsulated translation vector $(x, y, z)$ and the rotation vector containing the Euler angles $(\theta_x,\theta_y,\theta_z)$.
Then, for each of these two vectors, we determined the relative error 
(using the Euclidean norm) between the original and the interpolated one. 

\begin{figure}[tbh]
   \centering
   \includegraphics[width=\textwidth]{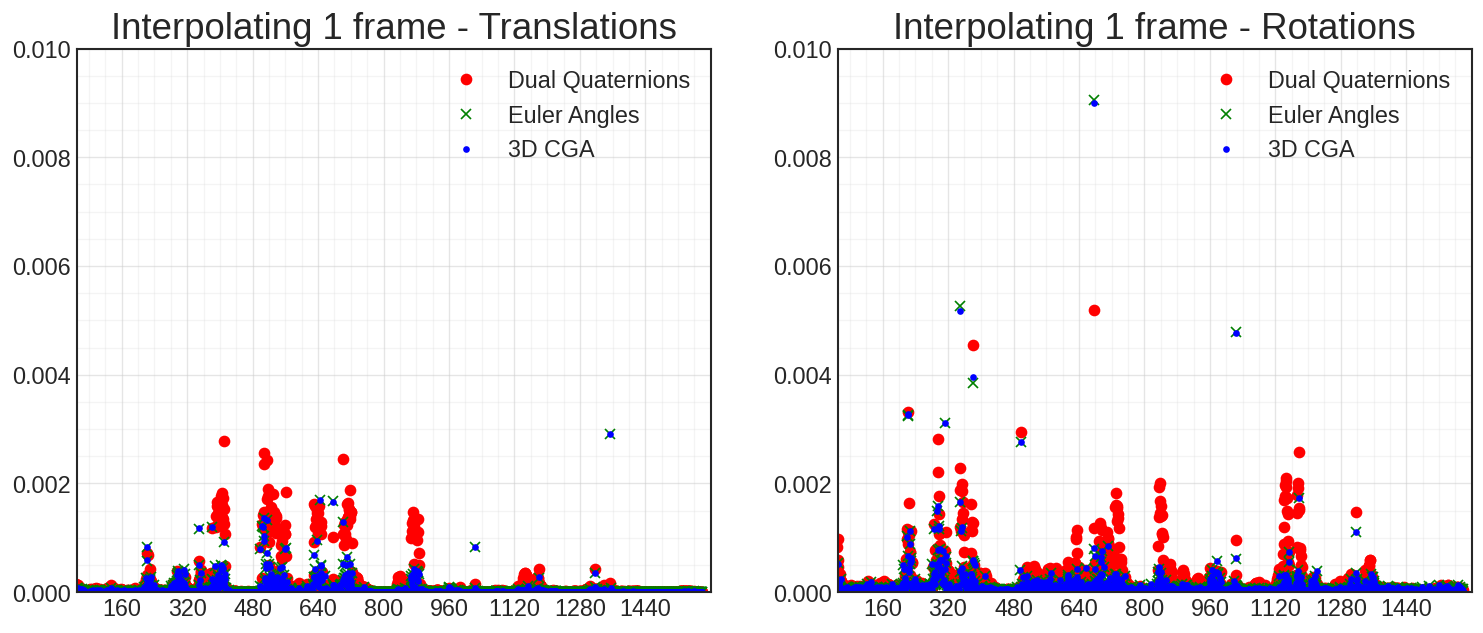}
   \caption{Results from Experiment A. The graph depicts the 
   relative error (vertical axis) that occurs if we replace the transitional (left)
   or the rotational data (right) of \textbf{one every two} frames 
   by interpolated data. This interpolated data depends on the 
   form used to represent the original displacement information; vectors 
   and Euler angles (green cross), dual quaternions (red dots) or 
   multivectors (blue dots). The horizontal axis refers
   the index of the frame for which the data are interpolated.}
   \label{fig:1_frame_interp}
 \end{figure}

\begin{figure}[tbh]
   \centering
   \includegraphics[width=\textwidth]{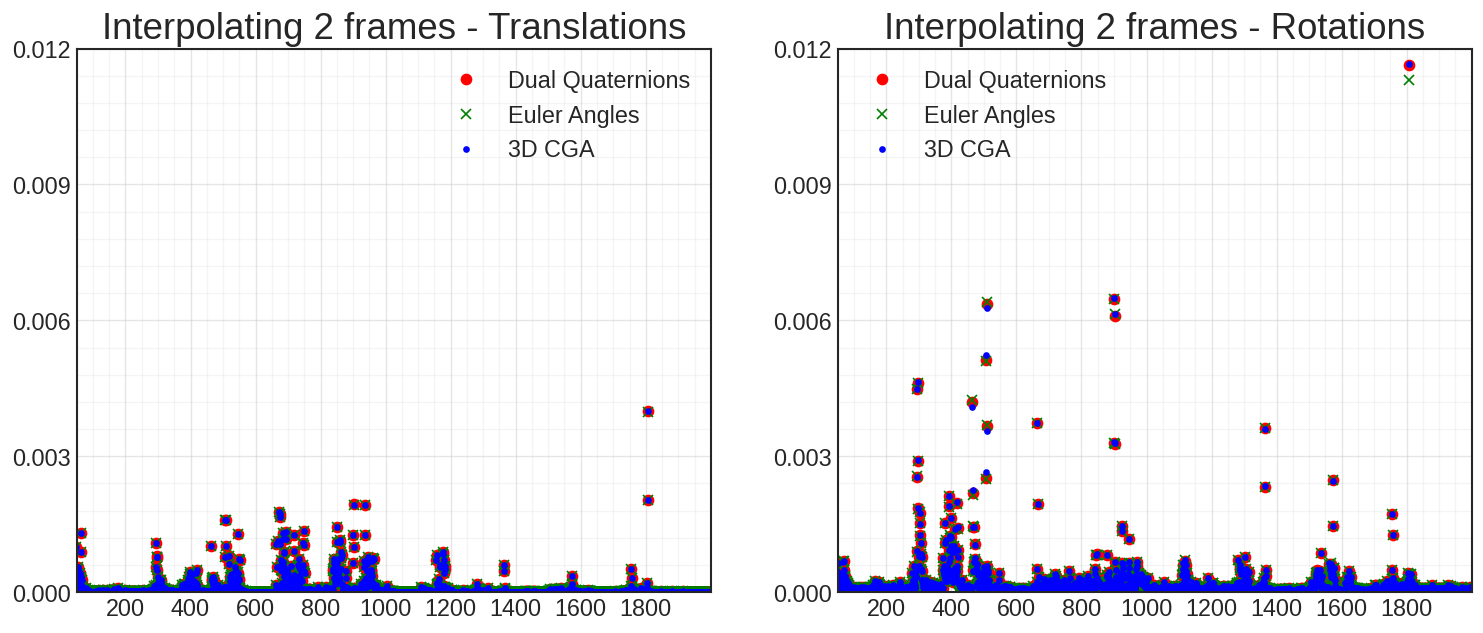}
   \caption{Results from Experiment B. The graph depicts the 
   relative error (vertical axis) that occurs if we replace the transitional (left)
   or the rotational data (right) of \textbf{one every three} frames 
   by interpolated data. This interpolated data depends on the 
   form used to represent the original displacement information; vectors 
   and Euler angles (green cross), dual quaternions (red dots) or 
   multivectors (blue dots).The horizontal axis refers
   the index of the frame for which the data are interpolated.}
   \label{fig:2_frame_interp}
 \end{figure}

Our findings are presented in Figure~\ref{fig:1_frame_interp} for 
the first experiment, and in Figure~\ref{fig:2_frame_interp} for the second 
one. In each experiment, we present the relative errors for the 
translational and rotational data separately. 
The results show that most displacement data have relative errors of less 
than $\mathbf{0.4\%}$ and $\mathbf{0.6\%}$ for interpolating one 
frame and two frames, respectively. 
In Tables~\ref{tbl:recorder_intep_trans} and \ref{tbl:recorder_intep_rot}, 
we present the mean relative errors that regard the translational and
rotational data respectively, for both experiments.
These quantitative findings,
along with the qualitative results of Section~\ref{sec:our_results} 
and \cite{kamarianakisCGI2021presentation},
prove that alternative, GA-based representation forms 
for displacement data are suitable for data recording as well
data transmission over networks for multi-player VR sessions.

Of course, one could try similar experiments and keep, e.g., the data of one every $n$ frames, for some random $n$. As $n$ grows larger, we emulate the transmission of less ``actual'' data and the creation of more ``artificial''
interpolated frames. Inevitably, the results obtained, although they would 
remain quite similar in all methods, they would not qualitatively (e.g., when visualized) correspond to the actual movement of the object on the scene. For example, collisions or pass-through of objects in the scene would 
start to happen, or de-synchronization of objects that were supposed to 
move in a synchronous way could arise in such situations. The maximum $n$
at which such phenomenons start to appear depends strictly on the nature 
of VR-session and the ``pace'' of the recorded data, i.e., if  
user data corresponds to slow smooth movements, we can safely omit recording more data and still obtain close-to-the-original interpolated frames.

\section{Conclusions and Future Work} 
\label{sec:conclusions}

This work suggests the use of two alternative representation forms, namely 
dual-quaternions and multivectors, to encapsulate the displacement 
information of the users in the context of a shared, collaborative 
virtual environment. These forms, based on dual-quaternions and 
multivectors of 3D Projective or Conformal Geometric Algebra, require 
specific interpolation techniques that we present here. 
Crucially, we provide the way to transmute between these forms and 
the classic SoA representation, i.e., vectors and quaternions for
translation and rotation. The benefits of using the proposed 
forms is demonstrated for two major VR functionalities;
data transmission over the network and data recording. 

Regarding data transmission over the network, we apply our proposed 
methods in a modern game engine and provide convincing
evidence (see the video in \cite{kamarianakisCGI2021presentation}) 
that they outperform SoA methods, 
offering increased QoE, especially as the network quality, in 
terms of bandwidth, deteriorates. 

Regarding data recording, we introduce a novel 
method that describes the functionality and characteristics 
of an efficient VR recorder with replay capabilities. 
Our proposed representation forms can be used to effectively 
store 3D scene transformation data.
In the case that specific data is intentionally omitted to be 
recorded, for storage efficiency purposes, 
the use of the suggested interpolation 
methods will generate data ``close'' to the omitted, with 
insignificant relative error.


Regarding future work, the results of our proposed interpolation 
algorithms can be further improved 
by using optimized C\# Geometric Algebra bindings to efficiently
perform complex operations such as dual-quaternion or multivector 
SLERP. Moreover, by using the data logged with the VR Recorder, 
we plan to create (a) an assessment tool that evaluates how the users perform on the given tasks,(b) intelligent agents that are able to 
complete tasks on their own and (c) a no-code VR authoring tool, 
with which the environment designers can develop new training modules.

The assessment tool will be developed by training supervised learning 
algorithms with labelled action data, provided  by the 
simulation designer. For example, the designer will provide multiple session data labelled from 0 to 10, depending on whether he/she performed the action poorly (0) or perfectly (10). The intelligent agents will be trained using 
such experts' session data and the no-code VR authoring tool will fuse 
all recorded data and train machine learning algorithms in order 
to understand what task the designer wants to develop. An initial step 
can be found in \cite{mlCuts}, where the authors have developed a Convolutional Neural Network (CNN) which is able to evaluate cuts in a surgical virtual environment. 
The transformation data (translations and rotations) of the cuts are fed to the neural network, and scores of the cuts, labelled by the surgical simulation designer, are the classes of the classification problem. 
In a future version, we plan to utilize dual quaternions and 
multivectors as the input form of the neural networks and examine 
whether data in such algebras can achieve better training results. 

Also, inspired by the Quaternion Convolutional Neural Network (QCNN) \cite{QCNN, qcnn2}, we  aspire to formulate and develop the Dual Quaternion Convolutional Neural Network (DQCNN) and the Multivector Convolutional Neural Network (MCNN). Lastly, QCNN algorithms can also be deployed to address the 
classification problem (similar to the one posed in \cite{mlCuts}), especially when used for rotational data, while keeping CNN for translational data. 
We assume that the overall accuracy will increase, since the CNNs treat the transformation data as six unrelated features, while the QCNN processes the rotations as a hypercomplex number, and therefore takes into consideration their interrelationship information.


\section{Acknowledgments  } 
\label{sec:Acknowledgments}

This work was co‐financed by European Regional Development 
Fund of the European Union and Greek national funds through the 
Operational Program Competitiveness, Entrepreneurship and Innovation, 
under the call RESEARCH – CREATE - INNOVATE 
(project codes:T1EDK-01149 and T1EDK-01448). 
The project also received funding from the European Union’s
Horizon 2020 research and innovation programme under grant agreement 
No 871793.


\bibliographystyle{spmpsci}
\bibliography{references}

\end{document}